\documentstyle[12pt]{article}

\begin{document}

\renewcommand{\thepage}{}

\begin{titlepage}


\title{
\hfill
\parbox{4cm}{\normalsize KUNS-1353\\HE(TH)\ 95/12 \\
hep-ph/9508269}\\
\vspace{5ex}
A SUSY SO(10) GUT with an Intermediate Scale
\vspace{5ex}}
\author{Joe Sato
   \thanks{e-mail address: {\tt joe@gauge.scphys.kyoto-u.ac.jp}}\\
{\it Department of Physics, Kyoto University,
      Kyoto 606-01, Japan}}
\date{\today}

\maketitle

\begin{abstract}
We examine a superpotential for an SO(10) GUT
and show that if the parameters of the superpotential
are in a certain region, the SO(10) GUT has
an intermediate symmetry
${\rm SU(2)}_L \otimes {\rm SU(2)}_R \otimes {\rm SU(3)}_C \otimes
{\rm U(1)}_{B-L}$ which breaks down to the
group of the Standard Model at an intermediate scale $10^{10-12}$ GeV.
In the model
by the breakdown of the symmetry right-handed neutrinos
acquire mass of the intermediate scale through a renormalizable
Yukawa coupling.
\end{abstract}

\end{titlepage}

\newpage
\renewcommand{\thepage}{\arabic{page}}

\section{Introduction}
When we construct a Grand Unified Theory(GUT) based on SO(10) \cite{so10},
in general, we have singlet fermions under the Standard Model(SM)
-what we call right-handed neutrino. Under the SM right-handed
neutrinos can have Majorana masses because they are singlet.
Then the scale of the right-handed neutrinos($\equiv M_{\nu_R}$)
is expected
to be a scale below which the SM is realized.

It is well known that in the Minimal Supersymmetric
Standard Model (MSSM) the present experimental values
of gauge couplings are successfully unified at a unification scale
$M_U \simeq 10^{16}$GeV \cite{Amal}. This fact implies
that if we would like to consider the gauge unification,
it is favorable that the symmetry of the GUT breaks down
to that of the SM at the unification scale.
In this case the scale of the right-handed neutrinos $M_{\nu_R}$ is
expected to be the unification scale $M_U$. This means also
there is no intermediate scale between
the Supersymmetry(SUSY) breaking scale and the unification scale.

On the other hand
it is said that $M_{\nu_R} \sim 10^{10-12} {\rm GeV}$\cite{yana}.
The experimental data on the deficit
of the solar neutrino can be explained by
the Mikheyev-Smirnov-Wolfenstein(MSW) solution \cite{MSW}.
According to one of the MSW solutions,
the mass of the muon neutrino seems to be
$m_{ \nu_\mu}\simeq10^{-3}$ eV.
Such a small mass can be led by the seesaw
mechanism \cite{seesaw}: A muon neutrino can
acquire a mass of O($10^{-3}$) eV
if the Majorana mass of the right-handed muon neutrino
is about $10^{12}$ GeV.

Then how can the right-handed neutrinos
acquire mass of  about $10^{12}$ GeV?
It was our question in our previous paper \cite{BST},
because if we take the prediction of the MSSM serious,
$M_{\nu_R}$ is expected to be $M_U \simeq 10^{16}$ GeV.
Our point of view was that it is more natural to consider
that one energy scale corresponds to a dynamical phenomenon,
for instance a symmetry breaking.
Mass is given by a renormalizable coupling
is also the crucial point of our view.
This idea is consistent with the survival hypothesis.
Thus we were led to a possibility that a certain group breaks
down to the SM group at the intermediate scale at which right-handed
neutrinos gain mass through a {\it renormalizable coupling}.

In the previous work
we have searched possibilities to construct such a SUSY
SO(10) GUT with an intermediate symmetry.
We have a possibility to construct a SUSY SO(10) GUT with an intermediate
symmetry ${\rm SU(2)}_L\otimes{\rm SU(2)}_R\otimes
{\rm U(1)}_{B-L}\otimes {\rm SU(3)}_C$ ($\equiv G_{2231}$)
\footnote{We use a notation $G_{lmn...}$ to represent
 ${\rm SU(l)}\otimes{\rm SU(m)}\otimes{\rm SU(n)}...$.If $l=1$,
it means U(1).}
which breaks down to the SM group at an intermediate
scale $M_{\nu_R} \sim 10^{10-12} {\rm GeV}$ where
a right-handed neutrino gains mass.

In such a scenario, as we showed in the previous work,
to make the model consistent with the gauge unification
we have to introduce several multiplets at the intermediate region
between the GUT scale and the intermediate scale,
in addition to ordinary matters, three generations of quarks and
leptons and a pair of so-called Higgs doublets.

Although we showed a possibility to construct
a SUSY SO(10) GUT with an intermediate symmetry $G_{2231}$
it is not trivial whether
it is actually possible to construct such a GUT
since there are many extra fields in the intermediate region.
We did not show the superpotential for the theory explicitly
which can realize such a scenario that we have suggested
in ref.\cite{BST}.

The purpose of this paper is to show an explicit form of
a superpotential for a SUSY SO(10) GUT to construct a
SUSY SO(10) GUT whose symmetry breaks
down to $G_{2231}$ at a GUT scale $M_U$ and $G_{2231}$
breaks down to the SM symmetry at the intermediate scale
$M_{\nu_R}$.

We give the scenario and the model briefly in sect. 2
where we give a candidate for the matter content in the intermediate
region (the spectrum (\ref{eq:spectrum})).
Then in sect. 3
we show the most general form of the superpotential
and a symmetry breaking condition
as preparation for our analysis.
In sect. 4 first we calculate parameters of the theory,
namely parameters appearing in the superpotential, which produce
the spectrum (\ref{eq:spectrum})
at the intermediate region. Then
we show the exact parameters which realize the MSSM
below $M_{\nu_R}$. Finally (in sect. 5)
we give a summary and discussion.

\section{Scenario and Model}

\subsection{Scenario}

We construct a SUSY SO(10) GUT whose symmetry breaks
down to $G_{2231}$ at a GUT scale $M_U$ and $G_{2231}$
breaks down to the SM symmetry at the intermediate scale
$M_{\nu_R}$. When $G_{2231}$ breaks down to the SM symmetry
the right-handed neutrinos gain mass through
a {\it renormalizable Yukawa coupling}.

Let us first recapitulate the content of the previous work\cite{BST}.
To achieve the gauge unification in the scenario
we have to introduce a certain combination of multiplets.
Because in our model right-handed neutrinos
acquire mass of O($M_{\nu_R}$) via a renormalizable
Yukawa coupling by the symmetry breaking,
we have to introduce at least a pair of
(1,3,1,6) + h.c multiplet under $G_{2231}$.
We adopt the
normalization for $U(1)_{B-L}$, $T^{15}_4 = {\rm diag}(-1,-1,-1,3)$.
When we introduce  only (1,3,1,6) + h.c multiplet in addition to the
ordinary matter,
gauge couplings do not unify. Then we have to introduce
certain matter content under $G_{2231}$.

We found very many
candidates for matter content in the intermediate region
between the GUT scale and the intermediate scale
which lead the gauge unification.
Among them we showed two candidates for the matter content as
the simplest example.
In this article we use another candidate
which was not showed in the previous
paper. In the examples appearing in it
a (1,3,1,0) multiplet under $G_{2231}$ was not included.
In constructing a GUT following the idea,
however, we have to introduce a
(1,3,1,0) multiplet in the intermediate region.
The reason why we have to introduce a (1,3,1,0) multiplet
is stated in the appendix A.
Thus we have to use another candidate for matter content.

The matter content
other than quarks and leptons (including right-handed neutrinos),
which we assume
survive until $G_{2231}$ breaks down to the SM group
at the intermediate scale,
are given below.

\begin{eqnarray}
\begin{array}{ccccl}
(1,3,1,-6) & 1 & (1,3,1,6) & 1
& {\rm responsible\,for }\nu_R {\rm  mass}\\
(2,2,1,0) & 2 && & {\rm ordinary\, Higgs\, doublets}\\
(2,1,3,-1) & 1 & (2,1,\overline{3},1) & 1& \\
(2,1,1,3) & 1 & (2,1,1,-3) & 1 &\\
(1,3,1,0) & 1 &&&\\
(1,1,8,0) & 1 &&&
\end{array}
\label{eq:spectrum}
\end{eqnarray}
In this list,
for example, (1,3,1,-6) 1 stands
for that the representation
of the matter under $G_{2231}$ is (1,3,1,-6) and its number is one.
When we have the particle content
listed here in the intermediate region
the unified coupling $\alpha_U (M_U)$ is about 1/18
if we take the intermediate scale to be $10^{12}$ GeV.
As a candidate which contains (1,3,1,0), this candidate
leads the smallest unified coupling.

In our scenario, at the GUT scale $M_U$ where SO(10) breaks
down to $G_{2231}$ almost of all particles
have mass of O($M_U$) while the particles listed in
(\ref{eq:spectrum}) as well as quarks and leptons
are left massless. Then at the intermediate scale
where $G_{2231}$ breaks down to $G_{231}$ the SM group
all extra multiplets in  (\ref{eq:spectrum})
besides a pair of Higgs doublets
and right-handed neutrinos have mass of O($M_{\nu_R}$),
that is, they decouple from the spectrum. Thus below
$M_{\nu_R}$ the MSSM is realized.

\subsection{Model}

\subsubsection{Matter content}

To have multiplets (\ref{eq:spectrum}) and quarks/leptons
at the intermediate region we introduce
following multiplets of SO(10).

\begin{eqnarray}
\begin{array}{rcrl}
&&{\rm SO(10)}&G_{2231}\\
H&:&10&(2,2,1,0),...\\
A&:&45&(1,3,1,0),(1,1,8,0),...\\
\Phi&:&126&(1,3,1,-6),(2,2,1,0),...\\
\overline{\Phi}&:&\overline{126}&(1,3,1,6),(2,2,1,0),...\\
\Delta&:&210&(1,3,1,0),(1,1,8,0),...\cr
\Psi_{i=1\sim 4}&:&16&(2,1,3,-1),(2,1,1,3),{\rm quarks/leptons}\\
\overline{\Psi}&:&\overline{16}&(2,1,\overline{3},1),(2,1,1,-3),...\\
\end{array}
\label{eq:so10multiplets}
\end{eqnarray}
In this list numbers in the column of SO(10) means SO(10)
representation. In the last column we show what representation
in (\ref{eq:spectrum}) is contained in the corresponding SO(10)
multiplet.

By the requirement that the right-handed neutrinos get mass through
a renormalizable coupling, we introduce 126 and $\overline{126}$.
As a candidate of ordinary Higgs doublets 10 is introduced.
There are other candidates for ordinary Higgs doublets
in 126 and $\overline{126}$. Then the ordinary Higgs doublets
will be a mixture of these three.
To break SO(10) to the SM group
via $G_{2231}$, namely to have the intermediate symmetry
$G_{2231}$, we use 45 and 210\footnote{
Using only 210 it is impossible to break SO(10) to $G_{231}$
through $G_{2231}$ \cite{Lee}.
We can break SO(10) to the SM group via $G_{2231}$
using 45 + 54. In this case
if there is no multiplet which
contains (1,3,1,0) other than 45 (3,1,1,0)
is also massless. The reason is that
mass terms for (1,3,1,0) and (3,1,1,0) come from
the mass term of 45 and the vacuum expectation value of 54
through the coupling $45^2 54$
and because of D parity\cite{DP} they are same as each other's.
Thus we can get rid of the possibility of using 45+54.}.
These also contain (1,3,1,0) and (1,1,8,0).
4 16's and 1 $\overline{16}$ represent 4 generation +
1 anti-generation. The reason why we introduce a pair of
16 and $\overline{16}$ is that they contain
(2,1,3,-1) + h.c and (2,1,1,3) + h.c .

At this stage
the matter content (\ref{eq:so10multiplets}) is just a
candidate which may realize our scenario.

As we will see, we can write down the superpotential
with these matter which realize our idea.

\subsubsection{Singlets under the SM group}

In the SO(10) multiplets (\ref{eq:so10multiplets})
there are many singlets under the SM symmetry
(see appendix B for the meaning of subscripts 1,...,0):
\begin{eqnarray}
\begin{tabular}{lllr}
Field&:&Component&Little Group\cr
A&:&$a_{12+34+56}\equiv \alpha $&$G_{2231}$\cr
&:&$a_{78+90}\equiv \beta $&$G_{241}$\cr
$\Phi$&:&$\phi_{1-2i,3-4i,5-6i,7-8i,9-0i}
\equiv \phi$&SU(5)\cr
$\overline{\Phi}$&:&$\overline{\phi}_{1+2i,3+4i,5+6i,7+8i,9+0i}
\equiv \overline{\phi}$&SU(5)\cr
$\Delta$&:&$\delta_{7890}\equiv a$&$G_{224}$\cr
&&$\delta_{1234+3456+5612}\equiv b$&$G_{2231}$\cr
&&$\delta_{(12+34+56)(78+90)}\equiv c$&$G_{2311}$\cr
$\Psi_{i=1\sim 4}$&:&$\psi_{i=1\sim 4}$&SU(5)\cr
$\overline{\Psi}$&:&$\overline{\psi}$&SU(5)\cr
\end{tabular}
\label{eq:vevtable}
\end{eqnarray}
where $a, b,$... stand for vacuum expectation values (VEV)
of the corresponding fields. Little group means
a remaining symmetry
when only a corresponding component has a VEV.
For example, when only $a$ gets a VEV SO(10) breaks down to
$G_{224}$.

Among them $a, b$ and $\alpha$ are $G_{2231}$ singlets
and hence their order of magnitudes is expected to
be the GUT scale $M_U \sim 10^{16}$
GeV. By assumption that SO(10) breaks down to $G_{2231}$ at the GUT scale,
$b$ or $\alpha$ must be of order $M_U$.
Others must be of order at most $M_{\nu_R} \equiv M_U \epsilon$
by assumption
because they are not $G_{2231}$ singlets.
Also $\overline\phi$ is required to be of order $M_{\nu_R}$,
\begin{equation}
\overline\phi
 \sim M_{\nu_R} ( = M_U \epsilon)
\label{eq:Condphi}
\end{equation}
because it gives masses to the right-handed neutrinos.
Of course, as we will see, there are constraints among
VEVs in addition to the well known constraints - F-flat
and D-flat condition
because we require certain multiplets must have mass
of O($M_{\nu_R}$).

\section{Preparation}

\subsection{Superpotential}

With the multiplets (\ref{eq:so10multiplets})
the most general form
of the superpotential $W$ is written as
\begin{eqnarray}
W = W_{mass} + W_{int} + W_{\Psi} .
\label{eq:superpotential}
\end{eqnarray}

$W_{mass}$ consists of the most general bilinear terms:
\begin{eqnarray}
W_{mass} = {1\over 2}M_H H^2 + M_\Phi \overline{\Phi} \Phi
  +{1\over 2}M_\Delta \Delta^2+{1\over 2} M_A A^2
  + M_\Psi \overline{\Psi} \Psi_4 .
\label{eq:supermass}
\end{eqnarray}

We define only $\Psi_4$ has a mass term with $\overline{\Psi}$,
because
by a redefinition of $\Psi_4$ , namely by a rotation among
$\Psi_{i=1-4}$,
it is possible that only the new $\Psi_4$
has a mass term with $\overline\Psi$.

We require all mass parameters are O($M_U$) because
$M_U$ is the natural order for them.

$W_{int}$ has the most general interaction terms without
16 and $\overline{16}$:
\begin{eqnarray}
W_{int} &=& Y_{H\Phi\Delta}H\Phi\Delta
          + Y_{H\overline{\Phi}\Delta}H\overline{\Phi}\Delta
          + {1\over {3!}} Y_\Delta \Delta^3
          + Y_{\Phi\Delta}\overline{\Phi} \Delta \Phi
          +Y_{\Phi A}\overline{\Phi} A \Phi \nonumber\\
        & + &{1\over 2} Y_{\Delta A^2} A^2 \Delta
          +{1\over 2} Y_{\Delta^2A} A \Delta^2 .
\label{eq:superyukawa}
\end{eqnarray}

We require all Yukawa couplings are at most O(1).
More exactly, as an expansion parameter for the perturbation
we require they are at most O(1). As a expansion parameter for the
perturbation they appear with multiplied by a certain overall
factor.
The overall factors for each couplings are given in appendix B.3.

Finally, $W_\Psi$ represents the most general interaction terms
with 16 and $\overline{16}$.
\begin{equation}
\begin{array}{ccl}
\displaystyle
W_\Psi& = &
\displaystyle
\sum^4_{i=3} Y_{\Psi\Delta i} \overline{\Psi}\Delta\Psi_i
       + \sum^4_{i=2} Y_{\Psi Ai} \overline{\Psi} A \Psi_i
       + \sum_{ij} y_{ij} \Psi_i \Psi_j \overline\Phi
       + y' \overline{\Psi} \,\overline{\Psi} \Phi \\
      &
\displaystyle
+ &
\displaystyle
\sum_{ij} \tilde{y}_{ij} \Psi_i \Psi_j H
+ \tilde{y}' \overline{\Psi}\, \overline{\Psi} H .\hfill
\end{array}
\label{eq:superpsi}
\end{equation}

By the same reason that only $\Psi_4$ has a mass term with
$\overline\Psi$, only $\Psi_{3,4}$ have couplings with $\Delta$
and  only $\Psi_{2,3,4}$ have couplings with $A$.

To see in which direction the gauge group SO(10) can break down
we examine the D-term and the F-term conditions.

\subsection{D-flat condition}

To keep the SUSY all D-terms must be zero up to SUSY braking scale:
$$
\Phi^\dagger T^a_\Phi \Phi +
\overline\Phi^\dagger T^a_{\overline\Phi} \overline\Phi +
\sum_i \Psi^\dagger_i T^a_\Psi \Psi_i
+\overline\Psi^\dagger T^a_{\overline\Psi} \overline\Psi
+\Delta^\dagger T^a_\Delta \Delta + A^\dagger T^a_A A = 0 .
$$
Since the D-term for real representations automatically
vanishes \cite{HeMe,BuDeSa},

\begin{eqnarray}
2 (|\phi|^2 - |\overline\phi|^2) +
(\sum_{i=1}^4 |\psi_i|^2 - |\overline\psi|^2) = 0
\label{eq:Dphipsi}
\end{eqnarray}
must be satisfied. The factor 2 reflects the difference of
U(1) charge which corresponds to a broken generator.

Later we put $\psi_i$'s and $\psi$ zero. In this case
\begin{equation}
|\phi|^2 - |\overline\phi|^2 = 0 .
\label{eq:Dphi}
\end{equation}

\subsection{F-flat condition}

First we examine the F-flat condition for 16 and $\overline{16}$
with a mass term for $(1,2,1,-3) + {\rm h.c}$ component because
the singlet components of 16 and $\overline{16}$
are contained in it and therefore
there is a relation between the mass term and the F-flat condition.
By such an examination  we see
that $\psi_i$ and $\overline\psi$ should be zero though it is not a strict
reason for it.

The F-flat condition for 16 and $\overline{16}$ are as follows:
(See appendix B to know how to calculate Clebsch-Gordan (CG)
coefficient)

\begin{eqnarray}
{\partial W \over \partial \psi_1}=
2 \sum_{j=1}^4 y_{1j} \psi_j \overline{\phi}=0,
\label{eq:Fpsi1}
\end{eqnarray}

\begin{eqnarray}
{\partial W \over \partial \psi_2}=
2 \sum_{j=1}^4 y_{2j} \psi_j \overline{\phi}
-Y_{\Psi A2}(\sqrt{6} i \alpha + 2 i \beta) \overline\psi=0,
\label{eq:Fpsi2}
\end{eqnarray}

\begin{eqnarray}
\displaystyle
{\partial W \over \partial \psi_3} &
=&
\displaystyle
2 \sum_{j=1}^4 y_{3j} \psi_j \overline{\phi}
-Y_{\Psi A3}(\sqrt{6} i \alpha + 2 i \beta) \overline\psi
\nonumber \\
&-&
\displaystyle
Y_{\Psi\Delta 3}(2 \sqrt{6} a + 6 \sqrt{2} b + 12 c) \overline\psi
\label{eq:Fpsi3}
\\&
\displaystyle
=&
\displaystyle
0,\nonumber
\end{eqnarray}

\begin{eqnarray}
\displaystyle
{\partial W \over \partial \psi_4}&=&
\displaystyle
2 \sum_{j=1}^4 y_{4j} \psi_j \overline{\phi}
-Y_{\Psi A4}(\sqrt{6} i \alpha + 2 i \beta) \overline\psi
\nonumber \\
&-&
\displaystyle
Y_{\Psi\Delta 4}(2 \sqrt{6} a + 6 \sqrt{2} b + 12 c) \overline\psi
+M_\Psi \overline\psi
\label{eq:Fpsi4} \\
&=&0,\nonumber
\end{eqnarray}

\begin{eqnarray}
\displaystyle
{\partial W \over \partial \overline\psi}&=&
\displaystyle
2\, y'\overline{\psi} \phi+
\sum_{i=2}^4
- Y_{\Psi Ai}(\sqrt{6} i \alpha + 2 i \beta) \psi_i\nonumber \\
&-&
\displaystyle
 \sum_{j=3}^4
Y_{\Psi\Delta i}(2 \sqrt{6} a + 6 \sqrt{2} b + 12 c) \psi_i
+M_\Psi \psi_4
\label{eq:Fbarpsi} \\
& =& 0. \nonumber
\end{eqnarray}

By the way in the intermediate region where $G_{2231}$ is realized,
$\beta = c = 0$ and the mass term for (1,2,1,-3)+h.c is given by
\begin{equation}
{\partial^2 W \over {\partial \psi_i \partial \overline\psi}}
=\pmatrix{0\cr
-  \sqrt{6} i Y_{\Psi A2} \alpha\cr
-\sqrt{6} i Y_{\Psi A3} \alpha
- 2 \sqrt{6} Y_{\Psi\Delta 3} (a + \sqrt{3} b)\cr
-\sqrt{6} i Y_{\Psi A4} \alpha
- 2 \sqrt{6} Y_{\Psi\Delta 4} (a + \sqrt{3} b)
+M_\Psi
}.
\label{eq:m1213}
\end{equation}
If $\phi, \overline\phi, \psi_i, \overline\psi \sim O(\epsilon)$,
using
F-flat conditions (\ref{eq:Fpsi2})\, -\, (\ref{eq:Fpsi4}),
all of elements of the mass term for (1,2,1,-3)+h.c,
(\ref{eq:m1213}), are calculated to be of order $M_{\nu_R}$.
This, however, contradicts with
the mass spectrum (\ref{eq:spectrum}).
Though we may be able to  make the some elements of the mass term
O($M_U$), for example, by making
$\overline\psi$ O($\epsilon^2$)
(with an appropriate value of $\psi_i,\overline\phi \sim
O(\epsilon)$), we put $\psi_i$ and
$\overline\psi$ zero since what we try to do is to show a possibility
of SUSY SO(10) GUT with an intermediate scale
and to take $\psi_i = \overline\psi = 0$ as the solution of
the F-flat conditions for 16
and $\overline{16}$ is the easiest way of it.

Then other F-term conditions are as follows:

\begin{eqnarray}
{\partial W \over \partial a}=
24\,{\sqrt{2}}\,i Y_{\Delta^2A}\,\alpha b -
  {{ Y_{\Delta A^2}\,\beta^2 }\over {2\,\sqrt{6}} } +
  {{{ Y_\Delta}\,{c^2}}\over {12\,{\sqrt{6}}}} + { M_\Delta} \,a +
  {{{ Y_{\Phi\Delta}}\,{ \overline{\phi}}\phi}\over {10\,{\sqrt{6}}}}=0,
\label{eq:Fa}
\end{eqnarray}

\begin{eqnarray}
\displaystyle
{\partial W \over \partial b}&=&
24\,{\sqrt{2}}\,i\,{ Y_{\Delta^2A}}\,a { \alpha} -
  {{{ Y_{\Delta A^2}}\,{{{ \alpha}}^2}}\over {3\,{\sqrt{2}}}} +
  {{{ Y_\Delta}\,{b^2}}\over {18\,{\sqrt{2}}}} \nonumber \\
&+&
\displaystyle
  24\,{\sqrt{2}}\,i\,{ Y_{\Delta^2A}}\,{ \beta}\,c +
  {{{ Y_\Delta}\,{c^2}}\over {18\,{\sqrt{2}}}} + { M_\Delta}\,b +
  {{{ Y_{\Phi\Delta}}\,\phi\,{ \overline{\phi}}}\over
{10\,{\sqrt{2}}}}
\label{eq:Fb}\\
&=&0,\nonumber
\end{eqnarray}

\begin{eqnarray}
\displaystyle
{\partial W \over \partial c}&=&
-{{{ Y_{\Delta A^2}}\,{ \alpha}\,{ \beta}}\over {{\sqrt{6}}}} +
  24\,{\sqrt{2}}\,i\,{ Y_{\Delta^2A}}\,b\,{ \beta} +
  {{{ Y_\Delta}\,a c}\over {6\,{\sqrt{6}}}} \nonumber \\
&+&
 \displaystyle
  16\,{\sqrt{6}}\,i\,{ Y_{\Delta^2A}}\,{ \alpha}\,c +
  {{{ Y_\Delta}\,b\,c}\over {9\,{\sqrt{2}}}} + { M_\Delta} c+
  {{{ Y_{\Phi\Delta}}\,\phi\,{ \overline{\phi}}}\over {10}}
\label{eq:Fc}\\
&=&0,\nonumber
\end{eqnarray}

\begin{eqnarray}
\displaystyle
 {\partial W \over \partial \alpha}&=&
24\,{\sqrt{2}}\,i\,{ Y_{\Delta^2A}}\,a\,b -
  {{{\sqrt{2}}\,{ Y_{\Delta A^2}}\,{ \alpha}\,b}\over 3} -
  {{{ Y_{\Delta A^2}}\,{ \beta}\,c}\over {{\sqrt{6}}}}\nonumber\\ &+&
\displaystyle
   8\,{\sqrt{6}}\,i\,{ Y_{\Delta^2A}}\,{c^2} + { M_A}\,\alpha +
  {{{\sqrt{6}}\,{ Y_{\Phi A}}\,\phi\,{ \overline{\phi}}}\over 10}
\label{eq:Falpha}\\
& =&0,\nonumber
\end{eqnarray}

\begin{eqnarray}
{\partial W \over \partial \beta}=
-{{{ Y_{\Delta A^2}}\,a\,{ \beta}}\over {{\sqrt{6}}}} -
  {{{ Y_{\Delta A^2}}\,{ \alpha}\,c}\over {{\sqrt{6}}}} +
  24\,{\sqrt{2}}\,i\,{ Y_{\Delta^2A}}\,b\,c + M_A \,\beta +
  {{{ Y_{\Phi A}}\,\phi\,{ \overline{\phi}}}\over 5}=0,
\label{eq:Fbeta}
\end{eqnarray}

\begin{eqnarray}
{\partial W \over \partial \phi}=
\left( { Y_{\Phi A}}\,\left( {{{\sqrt{6}}\,{ \alpha}}\over 10} +
       {{{ \beta}}\over 5} \right)  +
    { Y_{\Phi\Delta}}\,
\left( {a\over {10\,{\sqrt{6}}}} + {b\over {10\,{\sqrt{2}}}} +
       {c\over {10}} \right)  + { M_\phi} \right) \,{ \overline{\phi}}
   =0.
\label{eq:Fphi}
\end{eqnarray}

\section{Analysis}

The purpose of this paper is to give the input parameters appearing
in the superpotential (\ref{eq:superpotential}). Though
VEVs listed in (\ref{eq:vevtable})
are functions of the input parameters
we will express them in the term of the VEVs since
we know the desirable values of the VEVs.

\subsection{First Step}
First we check whether it is possible to break $SO(10)$ down to
$G_{2231}$ consistently with the requirement that
the spectrum (\ref{eq:spectrum})
remains massless up to O($\epsilon$) $\sim$ O($M_{\nu_R}/M_U$).

\subsubsection{Multiplets under $G_{2231}$}

First we show what multiplets exist in the SO(10) multiplets
(\ref{eq:so10multiplets}).

\begin{eqnarray}
\begin{array}{lrrrr}
{\rm Multiplet\, under}\, G_{2231}&{\rm under SO(10),\, contained in}&
{\rm NG1}&{\rm NG2}\cr
\hline
(2,2,1,0)&10,126,\overline{126}&&\cr
(1,1,3,2)+{\rm h.c}&10,126,\overline{126}&&\cr
(3,1,1,0)&45,210&&\cr
(1,3,1,0)&45,210&&\tilde z\cr
(1,1,3,-4)+{\rm h.c}&45,210&x&\tilde x\cr
(1,1,8,0)&45,210&&\cr
(2,2,3,2)+{\rm h.c}&45,210&y&\tilde y\cr
(3,1,1,6)+{\rm h.c}&126+\overline{126}&&\cr
(3,1,3,2)+{\rm h.c}&126+\overline{126}&&\cr
(3,1,6,-2)+{\rm h.c}&126+\overline{126}&&\cr
(1,3,1,-6)+{\rm h.c}&126+\overline{126}&&\tilde z\cr
(1,3,\overline{3},-2)+{\rm h.c}&126+\overline{126}&&\tilde x\cr
(1,3,\overline 6,2)+{\rm h.c}&126+\overline{126}&&\cr
(2,2,3,-4)+{\rm h.c}&126,\overline{126}&&\tilde y\cr
(2,2,8,0)+{\rm h.c}&126,\overline{126}&&\cr
(3,1,3,-4)+{\rm h.c}&210&&\cr
(1,3,3,-4)+{\rm h.c}&210&&\tilde x\cr
(3,1,8,0)+{\rm h.c}&210&&\cr
(1,3,8,0)+{\rm h.c}&210&&\cr
(2,2,1,6)+{\rm h.c}&210&&\cr
(2,1,3,-1)+{\rm h.c}&16+\overline{16}&&\tilde y\cr
(1,2,\overline 3,1)+{\rm h.c}&16+\overline{16}&&\tilde x\cr
(2,1,1,3)+{\rm h.c}&16+\overline{16}&&\cr
(1,2,1,-3)+{\rm h.c}&16+\overline{16}&&\tilde z\cr
\end{array}
\label{eq:multiplets}
\end{eqnarray}

In this table NG1 means a Nambu-Goldstone (NG) mode associated with the
breakdown of SO(10) to $G_{2231}$.
An NG mode associated with the SO(10) breaking down to
the SM group $G_{231}$ is contained in a multiplet with $\tilde x,
\tilde y$ and $\tilde z$ in the column NG2. Under $G_{231}$,
certain components of the multiplets with
$\tilde x$ ($\tilde y, \tilde z$)
have same quantum number and mix with each other.
One of combinations of $\tilde x$ ($\tilde y, \tilde z$)
is massless which is swallowed by a gauge boson.

There are also singlets of $G_{2231}$ which we denote $a, b$
and $\alpha$.

\subsubsection{F-flat condition}

In the intermediate region $c ,\beta , \phi =0$.
And hence the F-term conditions (\ref{eq:Fa}) - (\ref{eq:Fphi})
are reduced to

\begin{eqnarray}
{\partial W \over \partial a}=
24\,i\,{\sqrt{2}}\,{ Y_{\Delta^2A}}\,{ \alpha}\,b
 + { M_\Delta} a = 0 ,
\label{eq:Fa2231}
\end{eqnarray}

\begin{eqnarray}
{\partial W \over \partial b}=
24\,i\,{\sqrt{2}}\,a\,{ Y_{\Delta^2A}}\,{ \alpha} -
  {{{ Y_{\Delta A^2}}\,{{{ \alpha}}^2}}\over {3\,{\sqrt{2}}}} +
  {{{ Y_\Delta}\,{b^2}}\over {18\,{\sqrt{2}}}} +
  { M_\Delta} b = 0,
\label{eq:Fb2231}
\end{eqnarray}

\begin{eqnarray}
{\partial W \over \partial \alpha}=
24\,i\,{\sqrt{2}}\,{ Y_{\Delta^2A}} a\,b -
  {{{\sqrt{2}}\,{ Y_{\Delta A^2}}\,{ \alpha}\,b}\over 3}
 + { M_A} { \alpha} = 0.
\label{eq:Falpha2231}
\end{eqnarray}

\subsubsection{Tuning of parameters}

{}From now on
as we stated at the top of this section,
we express the input parameters in the term of the VEVs.

Using the F-flat conditions
(\ref{eq:Fa2231}) and (\ref{eq:Falpha2231}),
$M_\Delta$ and $M_A$ are expressed as follows:

\begin{eqnarray}
{ M_{\Delta}} = M_\Delta (Y_{\Delta^2A} , a , b , \alpha) =
    {{-24\,{\sqrt{2}}\,i\,{ Y_{\Delta^2A}}\,{ \alpha}\,b}\over a},
\label{eq:Md1}
\end{eqnarray}

\begin{eqnarray}
M_A = M_A(Y_{\Delta^2A} , Y_{\Delta A^2} , a , b ,
\alpha) =
    {-72\,\sqrt{2}\,i\,Y_{\Delta^2A} a\,b +
           \sqrt{2}\,Y_{\Delta A^2}\,\alpha\,b
\over {3 \alpha}}.
\label{eq:Ma1}
\end{eqnarray}

There is an additional constraint which is obtained by
eliminating $M_\Delta$ from equations (\ref{eq:Fa2231}) and
(\ref{eq:Fb2231}):

\begin{eqnarray}
-24\,{\sqrt{2}}\,i\,Y_{\Delta^2A} a^2\,{ \alpha} +
  { {Y_{\Delta A^2}\,a\,\alpha^2}\over {3\,\sqrt{2}} } -
  {{Y_{\Delta}\,a\,b^2}\over {18\,\sqrt{2}}} +
  24\,\sqrt{2}\,i\,Y_{\Delta^2A}\,\alpha\,b^2
=0.
\end{eqnarray}

We can interpret that
this constraint with (\ref{eq:Md1}) and (\ref{eq:Ma1})
is equivalent with that determinant of the mass matrix for (1,1,3,-4)
($\equiv M(1,1,3,-4)$
an explicit form is given at appendix C)
vanishes because
(1,1,3,-4) is an NG mode and hence
when we substitute VEVs into the mass matrix for it
there must be one massless mode which mean the determinant
vanishes.

\begin{eqnarray}
\displaystyle
\left.
\begin{array}{cl}
&\displaystyle
\det M(1,1,3,-4)\\
\displaystyle
=&
\displaystyle
M_A\,M_\Delta +
  {{Y_\Delta \,M_A \,b}\over {18\,\sqrt{2}}} -
  {{Y_{\Delta A^2}\,M_\Delta\,b}\over {3\,\sqrt{2}}}\\
\displaystyle
+&
\displaystyle
1152\,Y^2_{\Delta^2A}\,a^2
 + 16\,i\,Y_{\Delta A^2}\,Y_{\Delta^2A}\,a \alpha -
  {{Y^2_{\Delta A^2}\,\alpha^2}\over 18} -
  {{Y_\Delta\, Y_{\Delta A^2}\,b^2}\over 108} \\
\displaystyle
=&
\displaystyle
0.
\end{array}
\right.
\label{eq:detMass1134}
\end{eqnarray}

Now we required that one (1,1,8,0) mode be massless and therefore
determinant of the mass matrix for it ($\equiv M(1,1,8,0)$)
should vanish.

\begin{eqnarray}
\left.
\begin{array}{cl}
&
\displaystyle
\det M(1,1,8,0)\\
=&
\displaystyle
M_A\,M_\Delta -
  {{Y_\Delta \,M_A \,b}\over {18\,\sqrt{2}}} +
  {{Y_{\Delta A^2}\,M_\Delta\,b}\over {3\,\sqrt{2}}}\\
\displaystyle
+&
\displaystyle
1152\,Y^2_{\Delta^2A}\,a^2
 + 16\,i\,Y_{\Delta A^2}\,Y_{\Delta^2A}\,a \alpha -
  {{Y^2_{\Delta A^2}\,\alpha^2}\over 18} -
  {{Y_\Delta\, Y_{\Delta A^2}\,b^2}\over 108} \\
\displaystyle
=&
\displaystyle
0.
\end{array}
\right.
\label{eq:detMass1180}
\end{eqnarray}

Combining (\ref{eq:detMass1134}) and (\ref{eq:detMass1180})
with substituting (\ref{eq:Md1}) and (\ref{eq:Ma1}),
we find

\begin{eqnarray}
{{-8\,i}\over 3}\,Y_\Delta\,Y_{\Delta^2A}\,a^2 +
  {{Y_\Delta \,Y_{\Delta A^2}\,a\,\alpha}\over 27} +
  16\,i\,Y_{\Delta A^2}\,Y_{\Delta^2A}\,\alpha^2 = 0
\label{eq:M113411801}\\
\left( \left(
(\ref{eq:detMass1134}) - (\ref{eq:detMass1180}) \right) a \alpha /b^2
\right)\nonumber
\end{eqnarray}
and
\begin{eqnarray}
2304\,Y^2_{\Delta^2A} a^3+
  32\,i\,Y_{\Delta A^2}\,Y_{\Delta^2A}\,a\,\alpha -
  {{Y^2_{\Delta A^2}\,a\,\alpha^2}\over 9} -
  {{Y_{\Delta}\,Y_{\Delta A^2}\,a\,b^2}\over 54}\nonumber\\ -
  2304\,Y^2_{\Delta^2A}\,a\,b^2 -
  32\,i\,Y_{\Delta A^2}\,Y_{\Delta^2A}\,\alpha\,b^2 = 0
\label{eq:M113411802}\\
( ( (\ref{eq:detMass1134}) + (\ref{eq:detMass1180}) ) * a  ).
\nonumber
\end{eqnarray}

Solving a simultaneous equation (\ref{eq:M113411801})
and (\ref{eq:M113411802}) we get forms of
$Y_\Delta$ and $ Y_{\Delta A^2}$
as a function of $ Y_{\Delta^2A},a, b, \alpha $. Then
by substituting these expressions into
(\ref{eq:Md1}) and (\ref{eq:Ma1}) we find the following
three sets of solutions for
$M_\Delta$ , $M_A$ ,$Y_\Delta$ and $ Y_{\Delta A^2}$
as a function of $ Y_{\Delta^2A},a, b, \alpha $:

\vspace*{0.5cm}
$
\left.
\begin{array}{l}
M_{\Delta} \\
M_A \\
Y_\Delta \\
Y_{\Delta A^2}
\end{array}
\right\} =
$

solution 1::
\begin{eqnarray}
\left\{
\begin{array}{@{\,}l}
\displaystyle
     {{-24\,\sqrt{2}\,i\,Y_{\Delta^2A}\,\alpha\,b}\over a}\\
\displaystyle
    {{24\,\sqrt{2}\,i\,Y_{\Delta^2A}\,a\,b}\over \alpha}\\
\displaystyle
	{{-864\,i\,Y_{\Delta^2A}\,\alpha}\over a} \\
\displaystyle
     {{144\,i\,Y_{\Delta^2A} a}\over \alpha}
\end{array}
\right.
&\qquad\qquad\qquad\,&\qquad\qquad\qquad\qquad
\label{eq:solution1}
\end{eqnarray}

solution 2::
\begin{eqnarray}
\left\{
\begin{array}{@{\,}l}
\displaystyle
   {{-24\,\sqrt{2}\,i\,Y_{\Delta^2A}\,\alpha\,b} \over  a}\\
\displaystyle
   {{- 24 i Y_{\Delta^2A}\,b}\over  {{\sqrt{2}}\,a\,\alpha}}
         \left( - a^2 + 3\,{b^2} -
             \sqrt{a^4-10 a^2 b^2 + 9 b^4}
               \right)\\
\displaystyle
     {{-432\,i\,{ Y_{\Delta^2A}}\,{ \alpha}\,
          \left( -3\,{a^2} + 3\,{b^2} -
            {\sqrt{{a^4} - 10\,{a^2}\,{b^2} + 9\,{b^4}}} \right) }\over
       {\left( -{a^3} + 3\,a\,{b^2} -
          a\,{\sqrt{a^4 - 10\,a^2\,b^2 + 9\,b^4}}\right)}} ,\\
\displaystyle
     {- 36 i\, Y_{\Delta^2A} \over {a \alpha}}\,
         \left( - 3\,{a^2} + 3\,{b^2} -
             \sqrt{a^4-10 a^2 b^2 + 9 b^4}
               \right)\\
\end{array}
\right.
\label{eq:solution2}
\end{eqnarray}

solution 3::
\begin{eqnarray}
\left\{
\begin{array}{@{\,}l}
\displaystyle
     {{-24\,\sqrt{2}\,i\,Y_{\Delta^2A}\,\alpha\,b}\over a}\\
\displaystyle
\displaystyle
   {{- 24 i Y_{\Delta^2A}\,b}\over  {{\sqrt{2}}\,a\,\alpha}}
         \left( - a^2 + 3\,{b^2} +
             \sqrt{a^4-10 a^2 b^2 + 9 b^4}
               \right)\\
\displaystyle
     {{-432\,i\,{ Y_{\Delta^2A}}\,{ \alpha}\,
          \left( -3\,{a^2} + 3\,{b^2} +
            {\sqrt{{a^4} - 10\,{a^2}\,{b^2} + 9\,{b^4}}} \right) }\over
        {-{a^3} + 3\,a\,{b^2} +
          a\,{\sqrt{{a^4} - 10\,{a^2}\,{b^2} + 9\,{b^4}}}}}\\
\displaystyle
     {- 36 i\, Y_{\Delta^2A} \over {a \alpha}}\,
         \left( - 3\,{a^2} + 3\,{b^2} +
             \sqrt{a^4-10 a^2 b^2 + 9 b^4}
               \right)\\
\end{array}
\right.
\label{eq:solution3}
\end{eqnarray}
In other words, once $M_{\Delta}, M_A, Y_\Delta$ and $Y_{\Delta A^2}$
are set to be one of these solutions,
the VEVs of $a, b$ and $\alpha$ can be chosen at our will
and one of (1,1,8,0) mode becomes massless.

Because we require also that one (1,3,1,0) mode be massless,
determinant of the mass matrix for it ($\equiv M(1,3,1,0)$)
must be zero.

\begin{eqnarray}
\begin{array}{cl}
&
\displaystyle
\det M(1,3,1,0)\\
\displaystyle
 = &
\displaystyle
-{{Y_\Delta\,Y_{\Delta A^2}\,a^2 }\over {36}} -
  16\,i\,Y_{\Delta A^2}\,Y_{\Delta^2A}\,a\,\alpha -
  {{Y^2_{\Delta A^2}\,\alpha^2}\over 6} -
  {{Y_\Delta\,Y_{\Delta A^2}\,a\,b}\over {18\,\sqrt{3}}} \\
\displaystyle
+&
\displaystyle
  16\,{\sqrt{3}}\,i\,Y_{\Delta A^2}\,Y_{\Delta^2A}\,\alpha\,b
 +   1152\,Y^2_{\Delta^2A}\,b^2 +
  {{Y_\Delta\,M_A\,a}\over {6\,\sqrt{6}}} \\
\displaystyle
 +&
\displaystyle
  16\,\sqrt{6}\,i\,Y_{\Delta^2A}\,M_A\,\alpha+
  {{Y_\Delta\,M_A\,b}\over {9\,\sqrt{2}}} -
  {{Y_{\Delta A^2}\,M_\Delta\,a}\over {\sqrt{6}}} + M_A\,M_\Delta \\
\displaystyle
= &
\displaystyle
 0.
\end{array}
\label{eq:detMass1310}
\end{eqnarray}

Using (\ref{eq:detMass1310}) and
(\ref{eq:solution1})-(\ref{eq:solution3}),
we obtain a following equation
 which determine a relation between $a$ and $b$
corresponding to a set of above solutions respectively:

solution 1::
\begin{eqnarray*}
{a^2}\,
  \left( -3\,{a^2} + 7\,{\sqrt{3}}\,a\,b - 6\,{b^2} \right) = 0.
\end{eqnarray*}

solution 2::
\begin{eqnarray*}
-15\,{a^6} +
62\,{\sqrt{3}}\,{a^5}\,b +
       237\,{a^4}\,{b^2} - 280\,{\sqrt{3}}\,{a^3}\,{b^3} -
       249\,{a^2}\,{b^4} + 234\,{\sqrt{3}}\,a\,{b^5} + 27\,{b^6} \\
=\left(
       33\,{a^4}\, -
       50\,{\sqrt{3}}\,{a^3}\,b\, -
       78\,{a^2}\,{b^2}\, +
       78\,{\sqrt{3}}\,a\,{b^3}\, +
       9\,{b^4}\right)
{\sqrt{{a^4} - 10\,{a^2}\,{b^2} + 9\,{b^4}}}.
\end{eqnarray*}

solution 3::
\begin{eqnarray*}
15\,{a^6} - 62\,{\sqrt{3}}\,{a^5}\,b -
       237\,{a^4}\,{b^2} + 280\,{\sqrt{3}}\,{a^3}\,{b^3} +
       249\,{a^2}\,{b^4} - 234\,{\sqrt{3}}\,a\,{b^5} - 27\,{b^6} \\
=\left(
       33\,{a^4}\, -
       50\,{\sqrt{3}}\,{a^3}\,b\, -
       78\,{a^2}\,{b^2}\, +
       78\,{\sqrt{3}}\,a\,{b^3}\, +
       9\,{b^4}\right)
{\sqrt{{a^4} - 10\,{a^2}\,{b^2} + 9\,{b^4}}}.
\end{eqnarray*}

Numerically $a$ and $b$ must satisfy the following relation
respectively:

solution 1::
\begin{eqnarray}
a = \left\{
\begin{array}{@{\,}l}
{b}/{\sqrt{3}} ,\\
2\,{\sqrt{3}}\,b
\label{eq:ab1}
\end{array}
\right.
\end{eqnarray}

solution 2::
\begin{eqnarray}
a = \left\{
\begin{array}{@{\,}l}
   -0.987293\,b\\
   \left( -0.120361 -
         0.724007\,i \right) \,b\\
   \left( -0.120361 +
         0.724007\,i \right) \,b\\
   5.11238\,b
\label{eq:ab2}
\end{array}
\right.
\end{eqnarray}

solution 3::
\begin{eqnarray}
a = \left\{
\begin{array}{@{\,}l}
    -3.13416\,b\\
    -0.0643986\,b \\
    \left( 1.10047 -
         0.0616122\,i \right) \,b \\
    \left( 1.10047 +
         0.0616122\,i \right) \,b
\label{eq:ab3}
\end{array}
\right.
\end{eqnarray}

The solution 1 is the exact solution and the others are
exact up to O($\epsilon$).

In other words, if $a$ and $b$ satisfy these relations,
one (1,3,1,0) mode becomes massless.

Other requirements that two (2,2,1,0) modes,
one (1,3,1,-6) + h.c mode,
one (2,1,3,1) + h.c mode and one (2,1,1,-3) + h.c mode
be massless are easily satisfied
by tuning parameters such as $M_\Phi, M_H, Y_{H\Phi\Delta}$,
$Y_{H\overline{\Phi}\Delta}$ and so on.

To make (1,3,1,-6) + h.c mode massless,
from the mass term for it (see appendix C)
\begin{eqnarray}
{ M_\Phi}=
-\left({{{\sqrt{6}}\, Y_{\Phi A}\alpha}\over 10} +
  {{Y_{\Phi\Delta}\,a}\over {10\,{\sqrt{6}}}} +
  {{{ Y_{\Phi\Delta}}\,b}\over {10\,{\sqrt{2}}}}
\right).
\label{eq:Mphi1}
\end{eqnarray}

To make two (2,2,1,0) mode massless
we tune parameters  $M_H, M_\Phi, Y_{H\Phi\Delta}$
and $Y_{H\overline{\Phi}\Delta}$ so that the eigenvalue equation
for the mass matrix of (2,2,1,0)

\begin{eqnarray}
\displaystyle
{\lambda^3}
- { M_H}\,{\lambda^2}
+\left( -{{ Y^2_{H\overline{\Phi}\Delta} \,b^2
         }\over 10} - {{ Y^2_{H\Phi\Delta} \,b^2 }\over {10}} -
( {{{ Y_{\Phi\Delta}}\,b}\over {15\,{\sqrt{2}}}}
 + { M_{\Phi}} )^2 \,
\right) \,\lambda & \cr
-\left( {{{ Y_{\Phi\Delta}}\,b}\over {15\,{\sqrt{2}}}}
 + { M_{\Phi}} \right) \,
  \left(
M_H\,\left( {{{ Y_{\Phi\Delta}}\,b}\over {15\,{\sqrt{2}}}}
 + { M_{\Phi}} \right)
 + {{ Y_{H\overline{\Phi}\Delta A^2}
\,Y_{H\Phi\Delta}\,{b^2}} \over 5} \right) =0
\label{eq:eigeneq2210}
\end{eqnarray}
has two 0 solutions (exactly these two solutions may
have at most O($\epsilon$) solution)\footnote{Implicitly
it is assumed that the mass matrix for (2,2,1,0) is hermite, that is,
all parameters appearing in the mass matrix are real}.
The way of getting two zero eigenvalues is to tune
the zeroth and first terms of $\lambda$
zero. More exactly the zeroth term must be at most O($\epsilon^2$)
and the first term must be at most  O($\epsilon$).

To satisfy these constraint
\begin{eqnarray}
\begin{array}{c}
M_\Phi+{ {Y_{\Phi\Delta} b} \over {15 \sqrt{2}}} \sim O(\epsilon),
 \\
Y_{H\Phi\Delta} \sim Y_{H\overline\Phi\Delta} \sim O(\sqrt\epsilon).
\end{array}
\label{eq:M221001}
\end{eqnarray}

(\ref{eq:Mphi1}) and the first equation of (\ref{eq:M221001})
lead

\begin{eqnarray}
Y_{\Phi A}=-{ {\sqrt{3} a + b}\over {6 \sqrt{3} \alpha}}
Y_{\Phi\Delta}
\label{eq:Yda1}
\end{eqnarray}
up to O($\epsilon$).

Finally to make one (2,1,3,-1) + h.c mode and one (2,1,1,3) + h.c mode
massless, foe example, we can switch only couplings with subscript 4 on
and tune
\begin{eqnarray}
Y_{\psi\Delta} = {7 \over {16 \sqrt{3}}} i Y_{\psi A} \alpha/b,
\label{eq:ypd0}
\end{eqnarray}
\begin{eqnarray}
M_\Psi = -{3 \over {4 \sqrt{6}}} i Y_{\psi A} \alpha
-{7 \over {4 \sqrt{2}}} i Y_{\psi A} {a\over b}\alpha.
\label{eq:mp0}
\end{eqnarray}

\subsubsection{check mass matrices}

Now we know the necessary condition for the parameters
realizing the spectrum (\ref{eq:spectrum}).
Then we check all the mass matrices to examine
whether these parameters really
produce the spectrum (\ref{eq:spectrum}).

solution 1::

The solution 1 does not produce  the spectrum (\ref{eq:spectrum}),
because by substituting the solution 1 (\ref{eq:solution1})
into the mass matrix of (2,2,6,2) multiplet,
this multiplet is calculated to be massless.

solution 2::

First to see whether the solution 2 ,
(\ref{eq:solution2}) with a relation between $a$ and $b$
(\ref{eq:ab2}), is usable, we substitute (\ref{eq:ab2})
into (\ref{eq:solution2}).

\begin{eqnarray}
&\left.
\begin{array}{@{\,}l}
M_{\Delta} \\
  M_A
\end{array}
\right\}=
\left\{
\begin{array}{@{\,}l}
\left\{
\begin{array}{@{\,}l}
   24.3089\,i\,{\sqrt{2}}\,Y_{\Delta^2A}\,\alpha\\
   19.1441\,i\,{\sqrt{2}}\,Y_{\Delta^2A}
    b^2/ \alpha
\end{array}
\right.\\
\left\{
\begin{array}{@{\,}l}
    \left( 32.2574 +
    5.36258\,i \right) \,{\sqrt{2}}\,Y_{\Delta^2A}\,
      \alpha\\
    \left( -4.78842 +
    0.510831\,i \right)
    \,{\sqrt{2}}\,Y_{\Delta^2A} b^2 / \alpha
\end{array}
\right.\\
\left\{
\begin{array}{@{\,}l}
     \left( -32.2574 +
     5.36258\,i \right) \,{\sqrt{2}}\,Y_{\Delta^2A}\,
      \alpha \\
     \left( 4.78842 +
     0.510831\,i \right)
     \,{\sqrt{2}}\,Y_{\Delta^2A} b^2 / \alpha
\end{array}
\right.\\
\left\{
\begin{array}{@{\,}l}
     -4.69449\,i\,{\sqrt{2}}\,Y_{\Delta^2A}\,\alpha\\
     103.023\,i\,{\sqrt{2}}\,Y_{\Delta^2A} b^2
     / \alpha
\end{array}
\right.
\end{array}
\right.
\label{eq:sol2mdma}\\
&Y_{\Delta A^2} =
\left\{
\begin{array}{@{\,}l}
     {-13.6527\,i\, Y_{\Delta^2A}\,b} / \alpha \\
     {\left( 37.7632 -
       7.13352\,i \right) \,Y_{\Delta^2A}\,b} /
        \alpha \\
     {\left( -37.7632 -
       7.13352\,i \right) \,{ Y_{\Delta^2A}}\,b}
       / \alpha \\
     677.159\,i\,Y_{\Delta^2A}\,b/ \alpha
\end{array}
\right.
\label{eq:sol2ydaa}\\
& Y_\Delta =
\left\{
\begin{array}{@{\,}l}
     -104.016\,i\,Y_{\Delta^2A}\,\alpha / b\\
     \left( -1560.23 -
            131.862\,i \right) \,Y_{\Delta^2A}\,\alpha
         / b\\
     \left( 1560.23 -
            131.862\,i \right) \,Y_{\Delta^2A}\,\alpha
        / b \\
     -185.139\,i\,Y_{\Delta^2A}\,\alpha / b
\end{array}
\right.
\label{eq:sol2yd}
\end{eqnarray}

In each of these equations, four expressions correspond to
the four relations
between $a$ and $b$ in (\ref{eq:ab2}) respectively.

As we required that Yukawa couplings are not too big
(see the statement below (\ref{eq:superyukawa}))
only the first expression of the solution 2
is meaningful. This means that only the first relation
between $a$ and $b$ in (\ref{eq:ab2}) is meaningful.

By substituting (\ref{eq:solution2}) with the first equation of
(\ref{eq:ab2}) it is easy to check that all multiplets
other than those in (\ref{eq:spectrum}) have their mass of
O($M_U$) which spread around $M_U$ up to one order of magnitude
and multiplets in (\ref{eq:spectrum}) are massless.
Therefore this solution can be a solution of
our scenario.

solution3::

First we substitute (\ref{eq:ab3}) (relation between $a$ and $b$)
into (\ref{eq:solution3}) to see an explicit form of solution 3.

\begin{eqnarray}
&\left.
\begin{array}{@{\,}l}
M_{\Delta} \\
  M_A
\end{array}
\right\}=
\left\{
\begin{array}{@{\,}l}
\left\{
\begin{array}{@{\,}l}
     7.65756\,i\,{\sqrt{2}}\,Y_{\Delta^2A}\,\alpha\\
     -15.8066\,i\,{\sqrt{2}}\,Y_{\Delta^2A} b^2 /
     \alpha
\end{array}
\right.\\
\left\{
\begin{array}{@{\,}l}
    372.679\,i\,{\sqrt{2}}\,Y_{\Delta^2A}\,\alpha\\
    1115.98\,i\,{\sqrt{2}}\,Y_{\Delta^2A} b^2
    / \alpha
\end{array}
\right.\\
\left\{
\begin{array}{@{\,}l}
     \left( 1.21719 -
      21.7407\,i \right) \,{\sqrt{2}}\,Y_{\Delta^2A}\,
      \alpha\\
     \left( 17.3100 -
      22.7812\,i \right)
      \,{\sqrt{2}}\,Y_{\Delta^2A} b^2 /\alpha
\end{array}
\right.\\
\left\{
\begin{array}{@{\,}l}
     \left( -1.21719 -
     21.7407\,i \right) \,{\sqrt{2}}\,
     Y_{\Delta^2A}\,\alpha\\
     \left( -17.3100 -
     22.7812\,i \right) \,
      {\sqrt{2}}\,{ Y_{\Delta^2A}} b^2 / \alpha
\end{array}
\right.
\end{array}
\right.\\
&Y_{\Delta A^2} =
\left\{
\begin{array}{@{\,}l}
     -273.079\,i\,Y_{\Delta^2A}\,b / \alpha\\
     3343.29\,i\,Y_{\Delta^2A}\,b / \alpha\\
     \left( 56.3660 +
            10.8904\,i \right) \,Y_{\Delta^2A}\,b
       / \alpha \\
     \left( -56.3660 +
            10.8904\,i \right) \,Y_{\Delta^2A}\,b
        / \alpha
\end{array}
\right.\\
& Y_\Delta =
\left\{
\begin{array}{@{\,}l}
     793.766\,i\,Y_{\Delta^2A}\,\alpha / b \\
     6698.93\,i\, Y_{\Delta^2A}\,\alpha / b \\
     \left( 241.144 -
          102.803\,i \right) \,Y_{\Delta^2A}\,\alpha
          / b \\
     \left( -241.144 -
            102.803\,i \right) Y_{\Delta^2A}\, \alpha
         / b
\end{array}
\right.
\end{eqnarray}

In each of these equations, four expressions correspond to
the four relations
between $a$ and $b$ in (\ref{eq:ab3}) respectively.

By the same way as we picked
only the first expression up from four cases in solution 2,
the last two relations
between $a$ and $b$ in (\ref{eq:ab3}) are meaningful.

By substituting (\ref{eq:solution3}) with the third or fourth
equation of (\ref{eq:ab3}) it is easy to check that all multiplets
other than those in (\ref{eq:spectrum}) have their mass of
O($M_U$) which spread around $M_U$ up to one order of magnitude
and multiplets in (\ref{eq:spectrum}) are massless.
Therefore this solution can be a solution of
our scenario too.

\subsection{Second step}

In this section we find a parameter region
which produces our scenario exactly.

\subsubsection{Deviation from the previous solutions}

Because the accuracy of the previous calculation
is O($\epsilon$),
all parameters besides
$b, \alpha$ and $Y_{\Delta^2A}$
can deviate from the value which is obtained at the previous
section and therefore we can expand the deviation
in the power of $\epsilon$ as follows.

\begin{eqnarray}
\displaystyle
a = a_0 + \sum_{i=1} a_i \epsilon^i,
\label{eq:deva}
\end{eqnarray}

\begin{eqnarray}
\displaystyle
M_{\Delta} = M_{\Delta 0} + \sum_{i=1}  M_{\Delta i} \epsilon^i ,
\label{eq:devMd}
\end{eqnarray}

\begin{eqnarray}
M_A = M_{A 0} + \sum_{i=1}  M_{A i} \epsilon^i ,
\label{eq:devMa}
\end{eqnarray}

\begin{eqnarray}
Y_\Delta = Y_{\Delta 0}  + \sum_{i=1} Y_{\Delta i} \epsilon^i ,
\label{eq:devAd}
\end{eqnarray}

\begin{eqnarray}
Y_{\Delta A^2} = Y_{\Delta A^20}  + \sum_{i=1}
 Y_{\Delta A^2 i} \epsilon^i ,
\label{eq:devAdaa}
\end{eqnarray}

\begin{eqnarray}
\beta =  \sum_{i=1} \beta_i \epsilon^i ,
\label{eq:devbe}
\end{eqnarray}

\begin{eqnarray}
c =   \sum_{i=1} c_i \epsilon^i .
\label{eq:devc}
\end{eqnarray}

In these expressions, variables with subscript 0 stand for
those which are obtained in the previous section.

Substituting (\ref{eq:deva}) - (\ref{eq:devc}) into
the F-flat condition (\ref{eq:Fa}) - (\ref{eq:Fphi}),
we get following relations.

{}From (\ref{eq:Fa}), (\ref{eq:Fb}) and  (\ref{eq:Falpha}) we get

\begin{eqnarray}
 M_{\Delta 1} &=& -
{M_{\Delta 0}\over a_0} a_1,\nonumber\\
M_{A1} &=& {b^3 \over {9 \sqrt{2} \alpha^2}} Y_{\Delta 1}
+{{24 \sqrt{2} i Y_{\Delta^2A} b }\over \alpha}
(1+2 {b^2 \over a^2_0}) a_1,
\label{eq:deva1Md1Ma1Adaa1Ad1}\\
Y_{\Delta A^2 1}&=& (b^2/6 \alpha^2) Y_{\Delta 1}
+{{144 i Y_{\Delta^2A}}\over \alpha}(1+{b^2 \over a^2_0}) a_1.
\nonumber
\end{eqnarray}

We obtain the relation between $\beta_1$ and $c_1$ by
substituting (\ref{eq:deva}) - (\ref{eq:devc}) with
(\ref{eq:deva1Md1Ma1Adaa1Ad1}) into
(\ref{eq:Fc}) and (\ref{eq:Fbeta}) as follows:

First we note (\ref{eq:Fc}) and (\ref{eq:Fbeta})
can be rewritten

\begin{eqnarray}
M(1,3,1,0) \pmatrix{\beta\cr c}
=-{1 \over 10} \pmatrix{2 Y_{\Phi A}  \cr Y_{\Phi\Delta} }
\phi\overline\phi
\end{eqnarray}
and therefore

\begin{eqnarray}
\pmatrix{\beta\cr c}=
-{1 \over 10}
{\rm M}(1,3,1,0)^{-1}
\pmatrix{2 Y_{\Phi A}  \cr Y_{\Phi\Delta} }
\phi\overline\phi.
\label{eq:betaandc}
\end{eqnarray}
where M(1,3,1,0) is a mass matrix for (1,3,1,0) and by assumption
$\phi, \overline\phi \sim $ O($\epsilon$).

Let us decompose the inverse of M(1,3,1,0)

\begin{eqnarray}
{\rm M}(1,3,1,0)^{-1}=
\det\left({\rm M}(1,3,1,0)\right)^{-1}
\left({\rm A}+{\rm O}({\epsilon})\right).
\end{eqnarray}
Since by assumption there is one massless mode in (1,3,1,0)
up to O($\epsilon$), det(M(1,3,1,0)) $ \sim $ O($\epsilon$)
and the first row in A is parallel to the second row in A, that is
\begin{eqnarray}
{a_{11} \over a_{21}}= {a_{12} \over a_{22}}
\label{eq:aa}
\end{eqnarray}
where A $\equiv (a_{ij})$.

Then up to the leading order of $\epsilon$
\begin{equation}
\beta={a_{21} \over a_{11}} c
\label{eq:betac0}
\end{equation}
namely, as a exact relation
\begin{equation}
\beta_1={a_{21} \over a_{11}} c_1
\label{eq:betac1}
\end{equation}
is obtained.

To see this explicitly, we follow the above calculation
in the case of the first relation of solution 2.

\begin{eqnarray*}
\det({\rm M}(1,3,1,0))=
  (-26423.4  Y^2_{\Delta^2A} b\, a_1 + {{16.1727\,i
 Y_{\Delta^2A} Y_{\Delta 1} {b^3}}
\over \alpha})\,{ \epsilon}
+{\rm O} (\epsilon^2)
\end{eqnarray*}
as we expected the determinant is O($\epsilon$).

A is calculated to be
\begin{eqnarray*}
A=
\pmatrix{
72.3850\,i\,{ Y_{\Delta^2A}}\,{ \alpha}
,&
-39.5148\,i\,{ Y_{\Delta^2A}}\, b
\cr
 -39.5148\,i\,{ Y_{\Delta^2A}}\,b
,&
   {{21.5710\,i\,
{ Y_{\Delta^2A}}\,{b^2}} / {{ \alpha}}}
 }
\end{eqnarray*}
Apparently A satisfies (\ref{eq:aa}).

Then
\begin{eqnarray}
\beta_1=-1.83185 {\alpha \over b} c_1
\label{eq:betac}
\end{eqnarray}
is obtained.

\subsubsection{Determination of input parameters of the theory}

Though we can determine the parameters in the power of
$\epsilon$ order by order,
instead of doing so we will give the parameters of the theory
in a term of the VEVs
because the purpose of the paper is to find a parameter region
for the theory, $M$'s and $Y$'s,
which leads to the spectrum ({\ref{eq:spectrum}).
As we will see,
by the VEVs $a, b, c, \alpha$ and $\beta$
we can express the input parameters of the theory.

To do this, first we see the F-flat conditions (\ref{eq:Fa})
- (\ref{eq:Fbeta}). These equation can be rewritten

\begin{eqnarray}
C
\pmatrix{M_\Delta\cr
M_A\cr
Y_\Delta\cr
Y_{\Delta A^2}\cr
Y_{\Delta^2A}}=
-\pmatrix{1/(10 \sqrt{6}) Y_{\Phi\Delta}\cr
1/(10 \sqrt{2})  Y_{\Phi\Delta}\cr
1/10  Y_{\Phi\Delta}\cr
\sqrt{6} /10  Y_{\Phi A}\cr
 1/5  Y_{\Phi A}}
\phi\overline\phi
\end{eqnarray}
where

\begin{eqnarray}
C=\pmatrix{a,& 0,&{1\over{12 \sqrt{6}}} c^2,& - {1\over{2 \sqrt{6}}} \beta^2,&
24 \sqrt{2} i \alpha b \cr
 b,&0,& {1\over{18 \sqrt{2}}}  b^2 + {1\over{18 \sqrt{2}}} c^2
,&-{1\over{3 \sqrt{2}}}\alpha^2,&
24 \sqrt{2} i a \alpha + 24 \sqrt{2} i \beta c \cr
 c ,&0,& {1\over {6 \sqrt{6}}} a c + {1\over {9 \sqrt{2}}} b c,
&- {1\over{ \sqrt{6}}} \alpha \beta,
&16 \sqrt{6} i \alpha c + 24 \sqrt{2} i b \beta \cr 0,&
\alpha,&0,& - {\sqrt{2}\over 3} \alpha b - {1\over\sqrt{6}} \beta c,&
24 \sqrt{2} i a b + 8 \sqrt{6} i c^2 \cr
0,&\beta,&0,& - {1\over\sqrt{6}} \alpha c - {1\over\sqrt{6}} a \beta,&
24 \sqrt{2} i b c}
\end{eqnarray}

As we know from the previous argument that
$b,c$ and $\alpha$ can be chosen freely and
$a$ and $\beta$ are given by
\begin{eqnarray}
a = a_0 + a_1 \epsilon,\cr
\beta = \beta_1 \epsilon + \beta_2 \epsilon^2
\label{eq:constraintabeta}
\end{eqnarray}
where $a_0$ is given by the first equation of (\ref{eq:ab2})
or one of the last two equation of (\ref{eq:ab3})
and $\beta_1$ is given by (\ref{eq:betac1}). Note that
higher orders in (\ref{eq:deva}) and (\ref{eq:devbe})
can be absorbed into $a_1$ and $\beta_2$ respectively.

Then the input parameters are reduced to

\begin{eqnarray}
\pmatrix{M_\Delta\cr
M_A\cr
Y_\Delta\cr
Y_{\Delta A^2}\cr
Y_{\Delta^2A}}=
-C^{-1}
\pmatrix{1/(10 \sqrt{6}) Y_{\Phi\Delta}\cr
1/(10 \sqrt{2})  Y_{\Phi\Delta}\cr
1/10  Y_{\Phi\Delta}\cr
\sqrt{6} /10  Y_{\Phi A}\cr
 1/5  Y_{\Phi A}}
\phi\overline\phi .
\label{eq:parameters2}
\end{eqnarray}

For example, in the case of solution 2,
\begin{eqnarray*}
C^{-1}&=&(\det C)^{-1} C'\epsilon\\
\det C&=&(-3.76350\,i\,\alpha^2\,b^4\,\beta_2\,
   c_1-2.25347 i \alpha^3 b^2 a_1 c^2_1)
\,\epsilon^3 + {{{\rm O}({ \epsilon})}^4}
\end{eqnarray*}
\begin{eqnarray*}
C'&=&
\pmatrix{ 0 ,& 0,&
   -2.68018\,i\,\alpha^3\,b^3\,c_1 ,& 0,&
   -1.08826\,i\,\alpha^4\,{b^2}\,c_1\cr
 0 ,& 0,&
   -2.11074\,i\,\alpha\,{b^5}\, c_1 ,& 0,&
   -0.857040\,i\,\alpha^2\,{b^4}\,c_1\cr
 0 ,& 0,&
   8.10927\,i\,\alpha^3\,{b^2}\,c_1 ,& 0,&
   3.29268\,i\,\alpha^4\,b\,c_1 \cr
 0 ,& 0,&
  1.06439\,i\,\alpha\,{b^4}\,c_1 ,&0 ,&
  0.432184\,i\,\alpha^2\,{b^3}\,c_1 \cr
 0 ,& 0,&
   -0.0779620\,\alpha^2\,{b^3}\,c_1 ,&0 ,&
   -0.0316556\,\alpha^3\,{b^2}\, c_1} \\
&+&{\rm O}(\epsilon)
\end{eqnarray*}

{}From this equation it is easy to see that all parameters
are of order $\epsilon^0$ and they satisfy the first solution of
the solution 2.

Finally from (\ref{eq:Fphi}) $M_\Phi$ is determined:
\begin{equation}
{ M_\phi}=-
{ Y_{\Phi A}}\,\left( {{\sqrt{6}\,{ \alpha}}\over 10} +
       {{{ \beta}}\over 5} \right)  -
    { Y_{\Phi\Delta}}\,
\left( {a\over {10\,{\sqrt{6}}}} + {b\over {10\,{\sqrt{2}}}} +
       {c\over {10}} \right) .
\label{eq:Mphi2}
\end{equation}


\subsubsection{check mass matrices}

The multiplets in (\ref{eq:spectrum}) besides one (2,2,1,0)
must decouple at $M_{\nu_R}$,that is, they must acquire
mass of O$(M_{\nu_R})$.

{}From now on we check whether they have mass  of O$(M_{\nu_R})$.

First we note one (2,1,3,-1) + h.c and (2,1,1,3) + h.c
can have  mass  of O$(M_{\nu_R})$ by the following two reasons:
(1) Parameters $Y_{\psi\Delta}$ and $M_\Psi$ may deviate from the
value  given by (\ref{eq:ypd0}) and (\ref{eq:mp0})
respectively
\footnote{Though (2,1,3,-1) + h.c has a same
quantum number under the SM group as an NG mode
associated with the breakdown of SO(10)
the SM group (see table (\ref{eq:multiplets})),
it does not mix with others
because the VEV of $\psi=0$
and therefore
this NG mode does not consist of it.
(2,1,1,3) + h.c has a same quantum number as that of (2,2,1,0)
under the SM group but by the same reason they do not mix with
(2,2,1,0).
See the superpotential
(\ref{eq:superpotential}) - (\ref{eq:superpsi}). }.
(2) There exist couplings with $c$ and $\beta$.

Then we see the mass matrix for (2,2,1,0).
Under SM it has a quantum number (2,1,$\pm 1/2$).
(2,2,1,6) + h.c also includes the same component.
Then the mass matrix is
\begin{eqnarray}
M(2,1,\pm 1/2)&=&\pmatrix{
\tilde{M}_\Delta,&x,&y,&0\cr
x',&M_H,& u,& v\cr
0,&u,& 0,&, w-z\cr
y',&v,& w+z,&0}
\label{eq:massdoublets}
\end{eqnarray}
where
\begin{eqnarray}
\begin{array}{ccl}
\displaystyle
\tilde{M}_\Delta&=&{\rm M}(2,2,1,6)+{1\over12} Y_\Delta c + 24 i
Y_{A\Delta^2} \beta\\
x&=&-{1\over \sqrt{5}} Y_{H\overline{\Phi}\Delta} \overline\phi
 \sim {\rm O} (\epsilon^{3/2})\\
x'&=&-{1\over \sqrt{5}} Y_{H\Phi\Delta} \phi
 \sim {\rm O} (\epsilon^{3/2})\\
y&=&-{1\over 40} Y_{\Phi\Delta} \overline\phi
 \sim {\rm O} (\epsilon)\\
y'&=&-{1\over 40} Y_{\Phi\Delta} \phi
 \sim {\rm O}(\epsilon)\\
u&=&-{1\over \sqrt{10}} Y_{H\Phi\Delta}\,b +
{1\over {2 \sqrt{5}}} Y_{H\Phi\Delta}\,c \sim {\rm O}
(\sqrt{\epsilon}), \\
v&=& {1 \over \sqrt{10}}Y_{H\overline{\Phi}\Delta}\,b+
{1\over {2 \sqrt{5}}}  Y_{H\overline{\Phi}\Delta}\,c
\sim {\rm O} (\sqrt{\epsilon}),  \\
w&=& { M_\Phi}+
   {{{ Y_{\Phi\Delta}}\,b}\over {15\,{\sqrt{2}}}} \sim {\rm O}
(\epsilon) , \\
z&=&{{{ Y_{\Phi\Delta}}\,c}\over 30} +
   {{{ Y_{\Phi A}}\,\beta}\over 10} \sim {\rm O} (\epsilon) .
\end{array}
\label{eq:ahpd}
\end{eqnarray}
M(2,2,1,6) is given in the appendix C.
Orders of $x,y,..$ are followed from (\ref{eq:M221001})

Because one (2,1,$\pm 1/2$) multiplet remains massless
after $G_{2231}$ breaks down to the SM group
\begin{eqnarray}
\det\left(M(2,1,\pm 1/2)\right) =  \{ \tilde{M}_\Delta (z^2-w^2)+
y\, y'(w-z)\} M_H + 2 \tilde{M}_\Delta u v w ... =0,
\label{eq:detdoublets}
\end{eqnarray}
and hence $M_H$ is determined as follows:

\begin{eqnarray}
M_H= { {2 u v w} \over {w^2 - z^2}} + {\rm O} (\epsilon) .
\label{eq:mh}
\end{eqnarray}
In this case the higher order terms must be included
to have a pair of light Higgs doublets.

Next let us consider (1,1,8,0). This multiplet
becomes (1,8,0) under the SM group and therefore
it mixes with $T_{3R} = 0$ component of
(1,3,8,0) under the SM.
Then the mass matrix for (1,8,0) is represented as $3 \times 3$
matrix.
\begin{equation}
M(1,8,0)=\left(
\begin{array}{c|c}
M(1,1,8,0)&mixing\cr\hline
         mixing&M(1,3,8,0)
\end{array}
\right).
\label{eq:Mass180}
\end{equation}
After $G_{2231}$ breaks down to the SM group,
there is a correction of O($M_U \epsilon \sim M_{\nu_R}$)
to the mass
matrices  M(1,1,8,0) and M(1,3,8,0)
because parameters appearing in them
are different by O($\epsilon$)
from those calculated at the previous
section. It is directly calculated using (\ref{eq:parameters2})
(or equivalently (\ref{eq:deva}) - (\ref{eq:devAdaa})
and (\ref{eq:deva1Md1Ma1Adaa1Ad1}) ) that
one of the eigenvalues of  M(1,1,8,0) is of O($M_U$)
which has already suggested at the previous section and
the other is O($M_{\nu_R}$). As  M(1,3,8,0) is O($M_U$),
even though there is a correction of O($M_{\nu_R}$),
M(1,3,8,0) is still O($M_U$). Contributions of $c$ and $\beta$
to the mass matrix (\ref{eq:Mass180}) appear at mixing terms
between (1,1,8,0) and (1,3,8,0)\footnote{There is no contribution
of $c$ and $\beta$ to M(1,1,8,0) and M(1,3,8,0).
The reason is as follows. Under $G_{2231}$
 $c$ and $\beta$ are contained in (1,3,1,0). Because
(1,3,1,0)$(1,1,8,0)^2$  contains no singlet,  $c$ and $\beta$
do not couple to $(1,1,8,0)^2$. Though (1,3,1,0)$(1,3,8,0)^2$
can appear, as there is no three point coupling of $T_{3R}=0$
component of SU(2) triplet,  $c$ and $\beta$ do not couple to
$T_{3R}=0$ component of (1,3,8,0)}
 and they are of O($M_{\nu_R}$). Then M(1,8,0) takes the following
form
\begin{equation}
\left(
\begin{array}{ccc}
\rm O (M_U) &0&\rm O (M_U \epsilon) \\
0& O (M_U \epsilon)&\rm O (M_U \epsilon)\\
\rm O (M_U \epsilon)&\rm O (M_U \epsilon)&\rm O (M_U)
\end{array}
\right).
\end{equation}

Apparently two eigenvalues are of O($M_U$) and
the other is of O($M_{\nu_R}$). This fact suggests that
the lightest element of (1,1,8,0) under $G_{2231}$ decouples
at the scale $M_{\nu_R}$.

Finally we check the mass of (1,3,1,0) and (1,3,1,-6) + h.c.
Under the SM (1,3,1,0) is decomposed into one
neutral singlet
and a pair of charged singlet with
hypercharge $Y = \pm 1$. (1,3,1,,-6) + h.c becomes
two neutral singlets,  a pair of $Y = \pm 1$ and  a pair of
$Y = \pm 2$ singlets. Then $Y = \pm 1$
component of them  will
mix with each other.

Mass for  $Y = \pm 2$ component takes the following form
\begin{equation}
{ Y_{\Phi A}}\,\left( {{\sqrt{6}\,\alpha}\over 10} -
       {{{ \beta}}\over 5} \right)  +
    { Y_{\Phi\Delta}}\,
\left( {a\over {10\,{\sqrt{6}}}} + {b\over {10\,{\sqrt{2}}}} -
       {c\over {10}} \right)  + { M_\phi}
   = -{2 \over 5 }Y_{\Phi A}\beta - {1\over 5} Y_{\Phi\Delta} c
\end{equation}
where (\ref{eq:Mphi2}) is used.

{}From this equation obviously the $Y = \pm 2$ component
has a mass of O($M_{\nu_R}$).

Mass matrix of  $Y = \pm 1$ component is
\begin{equation}
\pmatrix{ -{{ Y_{\Delta A^2}\,a}\over {\sqrt{6}}} + M_A,&
   -{{{ Y_{\Delta A^2}}\,{ \alpha}}\over {{\sqrt{6}}}} +
    24\,i\,{\sqrt{2}}\,{ Y_{\Delta^2A}}\,b,&
   -{{Y_{\Phi A}\phi }\over 5} \cr
  -{{{ Y_{\Delta A^2}}\,{ \alpha}}\over {{\sqrt{6}}}} +
    24\,i\,{\sqrt{2}}\,{ Y_{\Delta^2A}}\,b,&
   {{Y_{\Delta}\,a}\over {6\,{\sqrt{6}}}} +
    16\,i\,{\sqrt{6}}\,{ Y_{\Delta^2A}}\,{ \alpha} +
    {{{ Y_{\Delta}}\,b}\over {9\,{\sqrt{2}}}} + { M_{\Delta}},&
   -{{Y_{\Phi\Delta}\phi }\over 10} \cr
  -{{Y_{\Phi A}\,\overline\phi}\over 5},&
   -{{Y_{\Phi\Delta}\,\overline\phi }\over 10},&
   -{{Y_{\Phi A}\,\beta}\over 5} -
    {{{ Y_{\Phi\Delta}}\,c}\over {10}} }.
\label{eq;MassY1}
\end{equation}

Since it is an NG mode associated with the breakdown of
$G_{2231}$ to $G_{231}$ there is one massless mode.
It is easy to see that this matrix has 0 eigenvalue because
${\rm 1st\ row} \times {\beta/\phi} + {\rm 2nd\ row}
\times c/\phi + {\rm 3rd\ row} = 0$ using the F-flat conditions
(\ref{eq:Fc}) and (\ref{eq:Fbeta}). It is also explicitly
calculated that one eigenvalue is of O($M_U$) and
the other is of O($M_{\nu_R}$).

\section{summary}

As we saw, by constructing the input parameters for the theory
 using (\ref{eq:parameters2}),
 (\ref{eq:Mphi2}), (\ref{eq:ahpd}) and (\ref{eq:mh})
from the desired values of VEVs $a, b, c, \alpha, \beta, \phi$
and $\overline\phi$ which satisfy (\ref{eq:Dphi}) and
(\ref{eq:constraintabeta}),
we can have particles (\ref{eq:spectrum}) in the intermediate region.
They decouple from the spectrum at $M_{\nu_R}$ except
a pair of what we call Higgs doublets.

It means that it is possible to construct a SUSY SO(10) GUT
with an intermediate scale
consistent with the gauge unification. It suggests also
that the right-handed neutrinos acquire mass through a
renormalizable coupling.
and it can be understood as a reflection of the breakdown of $G_{2231}$
to $G_{231}$

There are many variations for a SUSY SO(10) GUT with an
intermediate scale because there are many candidates
for the particle content which exist in the intermediate region
and we have many variations for content of SO(10) multiplets
which contain one of the candidates.

For example, we can replace (2,2,1,0) to (2,1,1,3) + h.c
in the spectrum (\ref{eq:spectrum}) and vise versa,
because their contribution to the running of the gauge coupling
relevant to $G_{231}$ is as same as that of each other.

When we remove one (2,2,1,0) from the spectrum (\ref{eq:spectrum})
and add one (2,1,1,3) + h.c to it,
by adding a pair of SO(10) multiplets $16 + \overline{16}$
which contains (2,1,1,3) + h.c under $G_{2231}$
we can have such a spectrum at the intermediate region.
At that time while we have to tune couplings relevant to
SO(10) multiplets $16 + \overline{16}$,
we can release the constraint (\ref{eq:M221001})
(or equivallently (\ref{eq:ahpd})).

Of course, there is a quite different type of content
for the candidates. Using them we can construct
quite a different SO(10) GUT with an intermediate scale.


Though the gauge unification by the MSSM is a very attractive idea,to
take into account a right handed-neutrino mass
we should consider a possibility of a GUT with an intermediate
symmetry.

\begin{center}
 ACKNOWLEDGEMENTS
\end{center}

The author wish to acknowledge T.~ Kugo, M.~Bando and
T.~Takahashi for valuable comments and discussion.

\appendix

\section{The reason why we need a multiplet (1,3,1,0)}

Here we show the reason why we need a multiplet (1,3,1,0)
in the intermediate region.

First we note that we required at least there is a pair of
multiplet (1,3,1,-6) + h.c ($\equiv \Phi + \overline\Phi$)
in the intermediate region \cite{BST}
and hence at this region in the superpotential
effectively there must be a term

\begin{equation}
W=M_\Phi \Phi \overline\Phi.
\label{eq:superapp1}
\end{equation}

Because we consider an SO(10) GUT the mass parameter
$M_\Phi$ is ,in general thought to be of O($M_U$).

In this case it is, however, impossible that
$\Phi$ acquires a VEV. Of course if we tune the parameter $M_\Phi$
be 0, as there is a flat direction in D-term, $\Phi$ can acquire
a VEV, but in this case there are two problems:

\noindent
(1) there is no way to determine a magnitude of the VEV of $\Phi$.

\noindent
(2) hypercharge Y $ = \pm 2$ component of
$\Phi$ cannot have mass\footnote{Note that
only an NG mode
can get a mass through D-term.
In general,
such a component corresponds to a massive gaugino.}.

Then we have to add other multiplets.
The easiest way to solve the problem (1)
is to add a singlet ($\equiv S$)
\footnote{Because we consider an SO(10) GUT, there are
several singlets though naturally their masses
are of O($M_U$).}. If there is a singlet
the superpotential will have a form

\begin{equation}
W=M_\Phi \Phi \overline\Phi + Y_{\Phi S} S \Phi \overline\Phi
+ {1 \over 2} M_S S^2 + {1 \over 3!} Y_S S^3
\end{equation}
and F-flat conditions are ($<\Phi> \equiv \phi, <S> \equiv s$)

\begin{eqnarray}
{{\partial W} \over {\partial \phi}} = ( M_\Phi + Y_{\Phi S} s)
\overline\phi =0 ,
\label{eq:Fphiapp}
\end{eqnarray}
\begin{eqnarray}
{{\partial W} \over {\partial s}} =  Y_{\Phi S} \phi \overline\phi
+M_S s + {1\over 2} Y_S s^2 .
\end{eqnarray}

Then VEVs are determined to
\begin{eqnarray}
s&=&-{M_\Phi \over Y_{\Phi S}} ,\\
 \phi \overline\phi &=& {M_S M_\Phi \over Y_{\Phi S}} -  {1\over 2} Y_S
({M_\Phi \over Y_{\Phi S}})^2.
\end{eqnarray}

Though as we mention below (\ref{eq:superapp1}) $M$'s are thought to be
of O($M_U$), we can give a VEV of O($M_{\nu_R}$) to $\Phi$
if coupling constants are fine-tuned while $s$ is of O($M_U$).

Unfortunately even after we add a singlet, the problem (2) is
not solved because the mass for Y=$\pm 2$ component is
\begin{equation}
M_\Phi+Y_S s=0
\end{equation}
according to the F-flat condition (\ref{eq:Fphiapp}). The reason why it is
still massless is
that no multiplet couples to $\Phi$ which
acquires a VEV of O($M_{\nu_R}$) and distinguishes
the component of a ${\rm SU(2)}_R$ triplet and hence
all component of $\Phi$
is still degenerate after  ${\rm SU(2)}_R$ breaking.

This means that to make Y=$\pm 2$ component decouple from the spectrum
after ${\rm SU(2)}_R$ breaking we have to make a multiplet couple
to $\Phi$ which will get a VEV of O($M_{\nu_R}$) and distinguishes
the component of a ${\rm SU(2)}_R$ triplet, that is,
a non-singlet.
It is easy to find
what non-singlet can couple to  $\Phi \overline\Phi$.
{}From  $\Phi \overline\Phi$ we have three representation:

\begin{eqnarray}
\begin{array}{c}
(1,1,1,0)\\
(1,3,1,0)\\
(1,5,1,0)
\end{array}
\end{eqnarray}

As ${\rm SU(2)}_R$ non-singlets are the latter two
and (1,5,1,0) is not contained in a relatively smaller
representation of SO(10),
we have to use (1,3,1,0). Since $T_{3R}=0$ component of a triplet is
an SM singlet it can get a VEV.

Since (1,3,1,0) is not a singlet under $G_{2231}$,
its VEV is at most of O($M_{\nu_R}$), while
because (1,3,1,0) gives a mass of O($M_{\nu_R}$)
to Y=$\pm 2$ component of $\Phi$, even if there are many
(1,3,1,0), one of their VEV must be of O($M_{\nu_R}$).
This implies that at least one of (1,3,1,0) must
have a mass of O($M_{\nu_R}$). In the following we will
see it explicitly.

First when there are also (1,3,1,0) multiplets
($\equiv B_i$) the superpotential takes a following form.
\begin{eqnarray}
W&=&M_\Phi \Phi \overline\Phi + Y_{\Phi S} S \Phi \overline\Phi
+ \sum_i Y_{i} B_i  \Phi \overline\Phi \nonumber\\
&+& {1 \over 2} M_S S^2 + {1 \over 3!} Y_S S^3
\label{eq:appsuperpotential}\\
&+& {1 \over 2} \sum_{i,j} (M_{ij}+ Y_{ij} S) B_i B_j
+{1\over 3!}\sum_{i,j,k} Y_{ijk} B_i B_j B_k\nonumber
\end{eqnarray}
and F-flat conditions are ($<B_i> \equiv \beta_i$)
\begin{eqnarray}
{{\partial W} \over {\partial \Phi}} = ( M_\Phi + Y_{\Phi S} s +
\sum_i Y_i \beta_i)
\overline\Phi =0 ,
\label{eq:Fphibeapp}
\end{eqnarray}
\begin{eqnarray}
{{\partial W} \over {\partial S}} =  Y_{\Phi S} \phi \overline\phi
+M_S s + {1\over 2} Y_S s^2  + \sum_{i,j} Y_{Sij} \beta_i \beta_j
=0,
\end{eqnarray}
\begin{eqnarray}
{{\partial W} \over {\partial B_i}} =  Y_i \phi \overline\phi
+ \sum_{i,j} (M_{ij}+ Y_{ij} s) \beta_i
=0.
\label{eq:Fbetaapp}
\end{eqnarray}

Note that there is no three point coupling of $T_3=0$ component
of SU(2) triplet and hence there is no affect of $Y_{ijk}$.

{}From (\ref{eq:Fbetaapp}) $\beta_i$ is calculated to
\begin{eqnarray}
\beta_i&=& -({\tilde M}^{-1})_{ij} a_j \phi \overline\phi ,\\
&&{\tilde M}_{ij} \equiv  (M_{ij}+ Y_{ij} s).\nonumber
\end{eqnarray}

By assumption $\phi \sim {\rm O}(M_{\nu_R})$
and as we mentioned one of $\beta_i$ also must be of O($M_{\nu_R}$).
These facts imply that in the above equation
${\tilde M}$ must have at least one eigenvalue of  O($M_{\nu_R}$).
Because ${\tilde M}$ is a mass matrix for (1,3,1,0) (see
(\ref{eq:appsuperpotential})),
it means that at least one of (1,3,1,0) must be massless
at the GUT scale.

In this case mass for Y$=\pm2$ is calculated
\begin{equation}
( M_\Phi + Y_{\Phi S} s -
\sum_i a_i \beta_i) = - 2 \sum_i a_i \beta_i \sim {\rm O}(M_{\nu_R})
\end{equation}
where (\ref{eq:Fphibeapp}) is used.
Apparently this component decouples at $M_{\nu_R}$, namely,
the problem (2) is solved.

\section{Construction of Representations}

In this section we briefly review how we construct representations
of subgroups contained in SO(10) representations
and give the rule for calculating CG coefficient appearing
in three point couplings. However, we do not mention about
an SO(10) spinor 16 because it is impossible
to understand the meaning of the indices for a spinor
in the same way of understanding that for an SO(10) vector 10
and essentially we do not need to handle them directly
in this paper.
To see how to handle an SO(10) spinor, see ref.\cite{Spin}.
When calculating CG coefficient relevant to a spinor
the gamma matrices for SO(10) constructed explicitly
in the reference are used.

\subsection{Meanings of Subscripts}

For SO(10) the fundamental representation
\footnote{Exactly in a mathematical term what fundamental
representation means is identity representation.}
is a 10 dimensional real vector

\begin{eqnarray*}
H = (H_i),\qquad i=1,...,10.
\end{eqnarray*}

It means when we construct a fundamental representation for SO(10)
we can use a following basis for it:

\begin{eqnarray}
H= h_i e_i,
\label{eq:vector}
\end{eqnarray}
where
\begin{eqnarray}
h_i = e^\dagger_i H,\
e_i \equiv \pmatrix{0\cr
                     \vdots \cr
                    0 \cr
                    1\cr
                    0\cr
                    \vdots\cr
                    0 }
\begin{array}{l}
\left.
\begin{array}{l}
\ \\
\ \\
\
\end{array}
\right.\\
\ \}i \ {\rm th\ component.}\\
\left.
\begin{array}{l}
\ \\
\ \\
\
\end{array}
\right.
\end{array}
\label{eq:base}
\end{eqnarray}
Here after in this appendix, repeated subscripts
are assumed to be contracted.

In this case index $i$ means nothing but SO(10) vector.

For our convenience we can attach an additional meaning to it.
SO(10) includes SU(5) $\otimes$ U(1) and
SO(6) $\otimes$ SO(4) $\simeq$ SU(4) $\otimes$
SU(2) $\otimes$ SU(2). Under them the fundamental representation 10
is decomposed into \cite{slan}
\begin{eqnarray*}
10=
\left\{
\begin{array}{ll}
5(2) + \overline{5}(-2) &\hbox{under  SU(5) $\otimes$ U(1)}\\
(6,1) + (1,4) &\hbox{under SO(6) $\otimes$ SO(4)} \\
(6,1,1) + (2,2,1) &\hbox{under SU(4) $\otimes$ SU(2) $\otimes$ SU(2)}
\end{array}
\right.
\end{eqnarray*}

Then we can add a meaning of, for example, SO(6) vector to
indices 1 to 6 and SO(4) vector to 7 to 10\footnote{
In the papers \cite{Lee,HeMe} they give a meaning of SO(6) vector
to indices 5 to 10 and that of SO(4) to 1 to 4}.
Here after 0 stands for 10. In other words
SO(6), an SO(10) subgroup, acts on the indices 1 to 6 and
SO(4) acts on 7 to 0.

We can add more meaning to indices of an SO(10) vector
by giving a meaning $5(2)$ representation under  SU(5) $\otimes$ U(1)
to $(1+2i, 3+4i, 5+6i, 7+8i, 9+0i)$ and its complex conjugate
to $(1-2i, 3-4i, 5-6i, 7-8i, 9-0i)$.

What $1+2i$ means is as follows.
When we construct a vector representation we can use
a basis $E_{a+bi}$ and its complex conjugate
$\overline{E}_{a-bi} \equiv \overline{E_{a+bi}}$
where $b=a+1$
and $a$ is an odd number other
than $e_i$ which is introduced at the top of this section.

\begin{eqnarray}
E_{a+bi}=
{1\over \sqrt{2}}
\left(
\begin{array}{c}
0\\
1\\
i\\
0
\end{array}
\right)
\begin{array}{l}
\\
\} a {\rm th}\\
\} b {\rm th}\\
\,
\end{array}
={1\over \sqrt{2}} e_a +{i\over\sqrt{2}} e_b
\label{eq:baseSU5}
\end{eqnarray}
where ${1\over\sqrt{2}}$ is a normalization factor to achieve
$E^\dagger_{a+bi} E_{a+bi} = 1$.

Then
\begin{eqnarray*}
H = h_i e_i = h_{a+bi} E_{a+bi} + h_{a-bi} \overline{E}_{a-bi}
\end{eqnarray*}
where
\begin{eqnarray}
h_{a+bi} = E^\dagger_{a+bi} H = {1\over\sqrt{2}}(h_a - h_b i)
\label{eq:SU5vector}
\end{eqnarray}
$h_{a+bi}$ is a component of a SU(5) vector and its U(1)
charge is 2.
As it is easily seen the component for an SO(10) vector depends on a
basis.

Because both SU(5) and SO(6) $\simeq$ SU(4) contain ${\rm SU(3)}_C$
we can add the meaning of SU(3) 3 and $\overline{3}$
to the SO(6) vector indices 1 to 6:
$(1+2i,3+4i,5+6i)$ is an SU(3) vector 3. By the same  way
we can add the meaning of SU(2) 2 and $\overline{2}$
to the SO(4) $\simeq$ SU(2) $\otimes$ SU(2)
vector indices 7 to 0: $(7+8i,9+0i)$ is an SU(2) vector 2.

As we will see later a higher representation
is represented as a tensor. By this construction when we consider
what representations a higher representation contains under,
for example, SO(10) subgroup SU(4), it is sufficient to
deal with indices 1 to 6. When considering SU(5) subgroup
we can deal with combinations of SO(10) subscripts $1+2i$ and so on.

\subsection{SO(10) Representations and
Representations of subgroups contained in SO(10) Representations}

The representations 45, 126 + $\overline{126}$ and 210
are formulated from the fundamental representation as
antisymmetric tensors of 2nd, 5th and 4th rank respectively.
By the characteristic of SO(10) 5th rank antisymmetric tensor
is decomposed into two parts, 126 and $\overline{126}$.
Using 10th rank antisymmetric $\epsilon$ tensor
($\equiv \varepsilon_{abcdeijklm}$) it is decomposed into
two eigenstates\cite{Spin}:
\begin{eqnarray}
\begin{array}{ccc}
{i\over {5!}} \varepsilon_{abcdeijklm} \Phi_{ijklm}& =& + \Phi_{abcde},\\
{i\over {5!}} \varepsilon_{abcdeijklm}
\overline{\Phi}_{ijklm}& =& - \overline{\Phi}_{abcde}.
\end{array}
\label{eq:eigen126}
\end{eqnarray}
What has a plus eigenvalue is defined to be 126
and the other is to be $\overline{126}$.

In the same way as an SO(10) vector 10 we can express these
representations using a component and a basis.
To express 45 ($\equiv A$) we can take a basis $e_{ij}$ as follows:
\begin{eqnarray}
A = a_{ij} e_{ij}
\label{eq:adjoint}
\end{eqnarray}
where
\begin{eqnarray}
a_{ij} = {\rm tr} A e_{ij},\
e_{ij} = \left
( (e_{ij})_{ab} \right) = {i\over\sqrt{2}}
( \delta_{ai}\delta_{bj}-\delta_{aj}\delta_{bi}).
\label{eq:base45}
\end{eqnarray}
$a_{ij}$ corresponds to a component of 45 representation.
In our notation subscripts $i,j$ for a component and a basis
satisfy that $i>j$.

In a similar manner 126 + $\overline{126}$ ($\equiv \Phi +
\overline{\Phi}$) is written as
\begin{eqnarray}
\Phi\, ({\rm or}\, \overline{\Phi}) = \phi_{ijklm} e_{ijklm}
\label{eq:onetwosix}
\end{eqnarray}
where $e_{ijklm}$ is an antisymmetric tensor and
only when a combination of indices coincide with
subscripts $\{ijklm\}$
it has a value $1/\sqrt{5!}$ or $-1/\sqrt{5!}$ . The sign is defined
to make $e_{ijklm}$ be antisymmetric.
Here $\{ijklm\}$ satisfies
$i>j>k>l>m$. Exactly, for $e_{ijklm}$
to be a basis of 126 (or $\overline{126}$) there is another constraint
for it as we explained at (\ref{eq:eigen126}),
though we do not touch the detail here. Then a component
of 126 is given by
\begin{eqnarray}
\phi_{ijklm} = \Phi_{abcde} (e_{ijklm})_{abcde}.
\label{eq:component126}
\end{eqnarray}
${1\over\sqrt{5!}}$ is a necessary normalization factor
to express a 126 representation by (\ref{eq:onetwosix}) and
(\ref{eq:component126}) similar to ${1\over\sqrt{2}}$ in
(\ref{eq:baseSU5}).

In the case of 210 a basis for it becomes 4th rank antisymmetric
tensor and its normalization is $1/\sqrt{4!}$.
Besides it 210 ($\equiv \Delta$) is represented in the same way:
\begin{eqnarray*}
\Delta = \delta_{ijkl} e_{ijkl}
\end{eqnarray*}
where
\begin{eqnarray*}
\delta_{ijkl} = \Delta_{abcd} (e_{ijkl})_{abcd}
\end{eqnarray*}
and $i>j>k>l$.

To construct a representation under subgroups
we use a linear combination of these basis in the same way
that when we extract a 5(2) of the subgroup SU(5) $\otimes$
U(1) from an SO(10) vector
we use a basis $E_{a+bi}$.

For example let us consider
$G_{231}$ singlets contained in 126 and $\overline{126}$.
They are SU(5) singlets. Then it is sufficient to deal with
SU(5) subscripts $1+2i$ and so on. By the quintality of
SU(5) the form of the basis of SU(5) singlets
in 126 and $\overline{126}$ are determined to be
$e_{1-2i,3-4i,5-6i,7-8i,9-0i},
e_{1+2i,3+4i,5+6i,7+8i,9+0i}$.
They are understood in the same way as $E_{1+2i}$,
(\ref{eq:baseSU5}):
\begin{eqnarray*}
e_{1-2i,3-4i,5-6i,7-8i,9-0i}={1\over\sqrt{10}}\left(
e_{13579}-i e_{23579}+...\right),
\end{eqnarray*}
where ${1\over\sqrt{10}}$ is an extra normalization factor to
achieve
\begin{eqnarray*}
(e_{1-2i,3-4i,5-6i,7-8i,9-0i})^*_{abcde}
(e_{1-2i,3-4i,5-6i,7-8i,9-0i})_{abcde}=1
\end{eqnarray*}
similar to ${1\over\sqrt{2}}$
in (\ref{eq:baseSU5}).

It is easily seen that
the former is a basis of 126 and the latter is that of
$\overline{126}$ by making  $\varepsilon_{abcdeijklm}$
acting on them or by counting U(1) charge\cite{slan}.
All other representation of subgroups contained in SO(10)
representations are constructed in a similar way.

\subsection{CG coefficient}
Using 10, 45, 126, $\overline{126}$ and 210
we have following SO(10) singlets\cite{slan}.
\begin{eqnarray*}
H\Phi\Delta, H\overline{\Phi}\Delta, \Delta^3,
 \overline{\Phi} \Delta \Phi,
 \overline{\Phi} A \Phi \nonumber,
 A^2 \Delta, A \Delta^2
\end{eqnarray*}

We can get singlets by contracting all indices of tensors:
\begin{eqnarray*}
H\Phi\Delta &\equiv& H_a \Phi_{abcde} \Delta_{bcde}\\
H\overline{\Phi}\Delta &\equiv&
 H_a \overline\Phi_{abcde} \Delta_{bcde}\\
 \Delta^3 &\equiv& \Delta_{abcd}\Delta_{cdef}\Delta_{efab}\\
\overline{\Phi} \Delta \Phi &\equiv&
 \overline{\Phi}_{abijk} \Delta_{abcd} \Phi_{cdijk}\\
\overline{\Phi} A \Phi &\equiv& \overline{\Phi}_{aijkl} A_{ab} \Phi_{bijkl}\\
A^2 \Delta &\equiv& A_{ab} A_{cd} \Delta_{abcd}\\
A \Delta^2 &\equiv&
 \varepsilon_{abcdefghij} A_{ab} \Delta_{cdef} \Delta_{ghij}
\end{eqnarray*}

In a term of components of the representations
\begin{eqnarray*}
H\Phi\Delta &=& {1\over\sqrt{5}}h_a \phi_{abcde} \delta_{bcde},\\
H\overline{\Phi}\Delta &=&  {1\over\sqrt{5}}
 h_a \overline\phi_{abcde} \delta_{bcde},\\
 \Delta^3 &=&  {1\over{6\sqrt{6}}}
 \delta_{abcd}\delta_{cdef}\delta_{efab},\\
\overline{\Phi} \Delta \Phi &=& {1\over{10\sqrt{6}}}
 \overline{\phi}_{abijk} \delta_{abcd} \phi_{cdijk},\\
\overline{\Phi} A \Phi &=&  {i\over{5\sqrt{2}}}
\overline{\Phi}_{aijkl} A_{ab} \Phi_{bijkl},\\
A^2 \Delta &=& -{1\over\sqrt{6}} a_{ab} a_{cd} \delta_{abcd},\\
A \Delta^2 &=&
 24\sqrt{2} i a_{ab} \delta_{cdef} \delta_{ghij}.
\end{eqnarray*}
where repeated subscripts are not summed and in the last equation
$abcdefghij$ are different from each other.

Then we rewrite the superpotential (\ref{eq:superpotential})
in a term of components, for example,
\begin{eqnarray*}
Y_\Delta \Delta^3 = {Y_\Delta\over{6\sqrt{6}}}
\delta_{abcd}\delta_{cdef}\delta_{efab}
\end{eqnarray*}
and so on.
Therefore for components that as an expansion
parameter for the perturbation Yukawa coupling = 1 means
$Y_\Delta = 6 \sqrt{6}$ and so on.

Of course, since a component of an irreducible representation
is a linear combination of these components, CG coefficient for
an irreducible representation is different from,
for example, ${1\over{6 \sqrt{6}}}$ in the case of $\Delta^3$.

For example let us calculate a CG coefficient for the singlet
$\beta$ contained in 45 and $a$ contained in 210 ( see the table
(\ref{eq:vevtable})).
They are contained in the form $A_{78+90}=\beta e_{78+90}$
and $\Delta_{7890}=a e_{7890}$ respectively. Then
\begin{eqnarray*}
A_{ab}A_{cd}\Delta_{abcd} &=& \beta^2 a  (e_{78+90})_{ab}
(e_{78+90})_{cd} (e_{7890})_{abcd}\\
&=&  \beta^2 a \left({i\over 2}\right)^2 {1\over\sqrt{4!}} 2! 2!
\times 2\\
&=& -{1\over\sqrt{6}} \beta^2 a.
\end{eqnarray*}
In the second line ${i\over 2}$ comes from an element of $e_{78+90}$
and ${1\over \sqrt{4!}}$ comes from an element of $e_{7890}$.
$2!$ comes from a summation between $\{ab\}$ and $\{cd\}$.
 $\{ab\}$ and $\{cd\}$ are $\{78\}$ or $\{90\}$.
The last factor 2 comes from an exchange of $\{78\}$ and $\{90\}$.

\section{
Mass matrices under $G_{2231}$ and their eigenvalue equations}

Under $G_{2231}$ the multiplets of our model have mass terms as follows.
They are listed following the order of the list (\ref{eq:multiplets}).
Full mass matrices are given with contributions from
$c, \beta, \phi$ and $\overline\phi$ after $G_{2231}$ breaks down to
$G_{231}$. But these contributions
are of order $M_{\nu_R} \sim M_U \epsilon$ and hence
if the mass eigenvalue is of O($M_U$), they are negligible
and we do not need to consider them.

 (2,2,1,0) multiplet;

\begin{eqnarray*}
M(2,2,1,0)=\pmatrix{
\displaystyle
{ M_H},&-{{{ Y_{H\Phi\Delta}}\,b}\over {{\sqrt{10}}}},&
   {{{ Y_{H\overline{\Phi}\Delta}}\,b}\over {{\sqrt{10}}}} \cr
   -{{{ Y_{H\Phi\Delta}}\,b}\over {{\sqrt{10}}}},& 0,&
   {{{ Y_{\Phi\Delta}}\,b}\over {15\,{\sqrt{2}}}} + { M_\Phi} \cr
  {{{ Y_{H\overline{\Phi}\Delta}}\,b}\over {{\sqrt{10}}}},&
   {{{ Y_{\Phi\Delta}}\,b}\over {15\,{\sqrt{2}}}} + { M_\Phi},&0
}
\end{eqnarray*}

 (1,1,3,-2) + h.c multiplet;

\begin{eqnarray*}
\displaystyle
M(1,1,3,2)=\pmatrix{
{ M_H},&
{{ Y_{H\Phi\Delta}\,\left( \sqrt{3} a - b \right) }\over
     \sqrt{30}},&{{ Y_{H\overline{\Phi}\Delta}\,
       \left(\sqrt{3} a + b \right) }\over \sqrt{30}} \cr
   {{{ Y_{H\Phi\Delta}}\,\left(\sqrt{3} a - b\right) }\over
     \sqrt{30}},&0
&{{{ Y_{\Phi A}}{ \alpha}}\over {5\,{\sqrt{6}}}} +
    { M_\Phi} \cr
  {{ Y_{H\overline{\Phi}\Delta}\,
       \left(\sqrt{3} a + b \right) }\over \sqrt{30}},&
   -{{{ Y_{\Phi A}}{ \alpha}}\over {5\,{\sqrt{6}}}} + { M_\Phi},&0
}
\end{eqnarray*}

 (3,1,1,0) + h.c multiplet;

\begin{eqnarray*}
\displaystyle
M(3,1,1,0)=\pmatrix{
 { M_A} + { {Y_{\Delta A^2}\,a} \over {\sqrt{6}} },&
   -{{{ Y_{\Delta A^2}} { \alpha}}\over {{\sqrt{6}}}} -
    24\,i\,{\sqrt{2}}\,{ Y_{\Delta^2 A}} b \cr
  -{{{ Y_{\Delta A^2}}\,{ \alpha}}\over {{\sqrt{6}}}} -
    24\,i\,{\sqrt{2}}\,{ Y_{\Delta^2 A}}\,b,&
   {{-{ Y_\Delta} a}\over {6\,{\sqrt{6}}}} -
    16\,i\,{\sqrt{6}}\,{ Y_{\Delta^2 A}}\,{ \alpha} +
    {{{ Y_\Delta}\,b}\over {9\,{\sqrt{2}}}} + { M_\Delta}
}
\end{eqnarray*}

 (1,3,1,0) multiplet;

\begin{eqnarray*}
\displaystyle
M(1,3,1,0)=\pmatrix{
 -{{{ Y_{\Delta A^2}} a}\over {{\sqrt{6}}}} + { M_A},&
   -{{{ Y_{\Delta A^2}}\,{ \alpha}}\over {{\sqrt{6}}}} +
    24\,i\,{\sqrt{2}}\,{ Y_{\Delta^2 A}}\,b \cr
   -{{{ Y_{\Delta A^2}}\,{ \alpha}}\over {{\sqrt{6}}}} +
    24\,i\,{\sqrt{2}}\,{ Y_{\Delta^2 A}}\,b,&
   {{{ Y_\Delta}a}\over {6\,{\sqrt{6}}}} +
    16\,i\,{\sqrt{6}}\,{ Y_{\Delta^2 A}}\,{ \alpha} +
    {{{ Y_\Delta}\,b}\over {9\,{\sqrt{2}}}} + { M_\Delta}
}
\end{eqnarray*}

 (1,1,3,-4) multiplet;

\begin{eqnarray*}
\displaystyle
M(1,1,3,-4)=\pmatrix{
 {{-{ Y_{\Delta A^2}}\,b}\over {3\,{\sqrt{2}}}} + { M_A},&
   24 {\sqrt{2}} i { Y_{\Delta^2 A}} a -
    {{{ Y_{\Delta A^2}}\,{ \alpha}}\over {3\,{\sqrt{2}}}}\cr
   24\,i\,{\sqrt{2}}\,{ Y_{\Delta^2 A}} a -
    {{{ Y_{\Delta A^2}}\,{ \alpha}}\over {3\,{\sqrt{2}}}},&
   {{{ Y_\Delta}\,b}\over {18\,{\sqrt{2}}}} + { M_\Delta}
}
\end{eqnarray*}

 (1,1,8,0) multiplet;

\begin{eqnarray*}
\displaystyle
M(1,1,8,0)=\pmatrix{
 {{{ Y_{\Delta A^2}}\,b}\over {3\,{\sqrt{2}}}} + { M_A},&
   24\,i\,{\sqrt{2}}\,{ Y_{\Delta^2 A}} a -
    {{{ Y_{\Delta A^2}}\,{ \alpha}}\over {3\,{\sqrt{2}}}} \cr
  24\,i\,{\sqrt{2}}\,{ Y_{\Delta^2 A}} a -
    {{{ Y_{\Delta A^2}}\,{ \alpha}}\over {3\,{\sqrt{2}}}},&
   -{{{ Y_\Delta}\,b  }\over {18\,{\sqrt{2}}}} + {
M_\Delta}
}
\end{eqnarray*}

 (2,2,3,2) + h.c multiplet;

\begin{eqnarray*}
\displaystyle
M(2,2,3,2)=\pmatrix{
  { M_A},&8 {\sqrt{6}} i { Y_{\Delta^2 A}}b,&
   -{{Y_{\Delta A^2}\,\alpha }\over 3} \cr
   8 {\sqrt{6}} i { Y_{\Delta^2 A}} b,&{ M_\Delta},&
   16 i {\sqrt{3}}{ Y_{\Delta^2 A}}{ \alpha} \cr
   -{{Y_{\Delta A^2}\,\alpha}\over 3},&
   16 {\sqrt{3}} i { Y_{\Delta^2 A}}\,{ \alpha},&
   {{{ Y_\Delta}\,b}\over {18\,{\sqrt{2}}}} + M_\Delta
}
\end{eqnarray*}

 (3,1,1,6) + h.c multiplet;

\begin{eqnarray*}
\displaystyle
M(3,1,1,6)=
-{{\sqrt{6}\,Y_{\Phi A} \,\alpha}\over 10}
- {{ Y_{\Phi\Delta}\,a}\over {10\,\sqrt{6}}} +
  {{{ Y_{\Phi\Delta}}\,b}\over {10\,{\sqrt{2}}}} + { M_\Phi}
\end{eqnarray*}

\vspace{1cm}

 (3,1,3,2) + h.c multiplet;

\begin{eqnarray*}
\displaystyle
M(3,1,3,2)=
-{{Y_{\Phi A} \alpha}\over {5\,{\sqrt{6}}}} -
  {{Y_{\Phi\Delta}\,a}\over {10\,\sqrt{6}}} +
  {{{ Y_{\Phi\Delta}}\,b}\over {30\,{\sqrt{2}}}} + { M_\Phi}
\end{eqnarray*}

\vspace{1cm}

 (3,1,6,-2) + h.c multiplet;

\begin{eqnarray*}
\displaystyle
M(3,1,6,-2)=
{{Y_{\Phi A}\,\alpha}\over {5\,\sqrt{6}}} -
  {{ Y_{\Phi\Delta}\,a}\over {10\,{\sqrt{6}}}} -
  {{{ Y_{\Phi\Delta}}\,b}\over {30\,{\sqrt{2}}}} + { M_\Phi}
\end{eqnarray*}

\vspace{1cm}

 (1,3,1,-6) + h.c multiplet;

\begin{eqnarray*}
\displaystyle
M(1,3,1,-6)=
{{{\sqrt{6}}\,Y_{\Phi A}\,\alpha}\over 10} +
  {{Y_{\Phi\Delta}\,a}\over {10\,\sqrt{6}}} +
  {{{ Y_{\Phi\Delta}}\,b}\over {10\,{\sqrt{2}}}} + { M_\Phi}
\end{eqnarray*}

\vspace{1cm}

 (1,3,3,-2) + h.c multiplet;

\begin{eqnarray*}
\displaystyle
M(1,3,3,-2)=
{{Y_{\Phi A}\,\alpha}\over {5\,{\sqrt{6}}}} +
  {{Y_{\Phi\Delta}\,a}\over {10\,{\sqrt{6}}}} +
  {{{ Y_{\Phi\Delta}}\,b}\over {30\,{\sqrt{2}}}} + { M_\Phi}
\end{eqnarray*}

\vspace{1cm}

 (1,3,6,2) + h.c multiplet;

\begin{eqnarray*}
\displaystyle
M(1,3,6,2)=
-{{Y_{\Phi A}\,\alpha}\over {5\,{\sqrt{6}}}} +
  {{Y_{\Phi\Delta}\,a}\over {10\,{\sqrt{6}}}} -
  {{{ Y_{\Phi\Delta}}\,b}\over {30\,{\sqrt{2}}}} + { M_\Phi}
\end{eqnarray*}

\vspace{1cm}

 (2,2,3,-4) + h.c multiplet;

\begin{eqnarray*}
\displaystyle
M(2,2,3,-4)=\pmatrix{
 {{\sqrt{6}\,Y_{\Phi A}\,\alpha}\over 15} +
    {{{ Y_{\Phi\Delta}}\,b}\over {30\,{\sqrt{2}}}} + { M_\Phi},&0 \cr
 0,&-{{\sqrt{6}\,
Y_{\Phi A}\,\alpha}\over 15} +
    {{{ Y_{\Phi\Delta}}\,b}\over {30\,{\sqrt{2}}}} + { M_\Phi}}
\end{eqnarray*}

\vspace{1cm}

 (2,2,8,0) multiplet;

\begin{eqnarray*}
\displaystyle
M(2,2,8,0)=
-{{{ Y_{\Phi\Delta}}\,b }\over
{30\,{\sqrt{2}}}} + { M_\Phi}
\end{eqnarray*}

\vspace{1cm}

 (3,1,3,-4) + h.c multiplet;

\begin{eqnarray*}
\displaystyle
M(3,1,3,-4)=
-{{Y_\Delta a}\over {6\,{\sqrt{6}}}} -
  8\,i\,{\sqrt{6}}\,{ Y_{\Delta^2 A}}\,{ \alpha} +
  {{{ Y_\Delta}\,b}\over {18\,{\sqrt{2}}}} + { M_\Delta}
\end{eqnarray*}

\vspace{1cm}

 (1,3,3,-4) + h.c multiplet;

\begin{eqnarray*}
\displaystyle
M(1,3,3,-4)=
{{Y_\Delta\,a}\over {6\,{\sqrt{6}}}} +
  8\,i\,{\sqrt{6}}\,{ Y_{\Delta^2 A}}\,{ \alpha} +
  {{{ Y_\Delta}\,b}\over {18\,{\sqrt{2}}}} + { M_\Delta}
\end{eqnarray*}

\vspace{1cm}

 (3,1,8,0) multiplet;

\begin{eqnarray*}
\displaystyle
M(3,1,8,0)=
-{{ Y_\Delta\,a }\over {6\,{\sqrt{6}}}} +
  8\,i\,{\sqrt{6}}\,{ Y_{\Delta^2 A}}\,{ \alpha} -
  {{{ Y_\Delta}\,b}\over {18\,{\sqrt{2}}}} + { M_\Delta}
\end{eqnarray*}

\vspace{1cm}

 (1,3,8,0) multiplet;

\begin{eqnarray*}
\displaystyle
M(1,3,8,0)=
{{Y_\Delta\,a}\over {6\,{\sqrt{6}}}} -
  8\,i\,{\sqrt{6}}\,{ Y_{\Delta^2 A}}\,{ \alpha} -
  {{{ Y_\Delta}\,b}\over {18\,{\sqrt{2}}}} + { M_\Delta}
\end{eqnarray*}

\vspace{1cm}

 (2,2,1,6) + h.c multiplet;

\begin{eqnarray*}
\displaystyle
M(2,2,1,6)=
{{{ Y_\Delta}\,b}\over {6\,{\sqrt{2}}}} + { M_\Delta}
\end{eqnarray*}

\vspace{1cm}

 (2,2,6,-2) + h.c multiplet;

\begin{eqnarray*}
\displaystyle
M(2,2,6,-2)=
-{{{ Y_\Delta}\,b }\over {18\,{\sqrt{2}}}} + M_\Delta
\end{eqnarray*}

 (2,1,3,-1) + h.c multiplet;

\begin{eqnarray*}
\displaystyle
M(2,1,3,-1)=
\pmatrix{0\cr
- {1 \over \sqrt{6}} i Y_{\psi_2A} \alpha\cr
-{1 \over \sqrt{6}} i Y_{\psi_3A} \alpha
+2 Y_{\psi_3\Delta} (\sqrt {6} a + \sqrt{2} b)\cr
-{1 \over \sqrt{6}} i Y_{\psi_4A} \alpha
+2 Y_{\psi_4\Delta} (\sqrt{6} a + \sqrt{2} b)
+M_\Psi
}
\end{eqnarray*}

 (1,2,$\overline 3$,1) + h.c multiplet;

\begin{eqnarray*}
\displaystyle
M(1,2,\overline 3,1)=
\pmatrix{0\cr
 {1 \over \sqrt{6}} i Y_{\Psi A2} \alpha\cr
{1 \over \sqrt{6}} i Y_{\Psi A3} \alpha
+2 Y_{\Psi\Delta 3} (- \sqrt {6} a + \sqrt{2} b)\cr
{1 \over \sqrt{6}} i Y_{\Psi A4} \alpha
+2 Y_{\Psi\Delta 4} (- \sqrt{6} a + \sqrt{2} b)
+M_\Psi
}
\end{eqnarray*}

 (2,1,1,3) + h.c multiplet;

\begin{eqnarray*}
\displaystyle
M(2,1,1,3)=
\pmatrix{0\cr
  \sqrt{6} i Y_{\Psi A2} \alpha\cr
\sqrt{6} i Y_{\Psi A3} \alpha
+ 2 \sqrt{6} Y_{\Psi\Delta 3} (a - \sqrt{3} b)\cr
\sqrt{6} i Y_{\Psi A4} \alpha +
 2 \sqrt{6} Y_{\Psi\Delta 4} (a - \sqrt{3} b)
+M_\Psi
}
\end{eqnarray*}

 (1,2,1,-3) + h.c multiplet;

\begin{eqnarray*}
\displaystyle
M(1,2,1,-3)=
\pmatrix{0\cr
-  \sqrt{6} i Y_{\Psi A2} \alpha\cr
-\sqrt{6} i Y_{\Psi A3} \alpha
- 2 \sqrt{6} Y_{\Psi\Delta 3} (a + \sqrt{3} b)\cr
-\sqrt{6} i Y_{\Psi A4} \alpha
- 2 \sqrt{6} Y_{\Psi \Delta4} (a + \sqrt{3} b)
+M_\Psi
}
\end{eqnarray*}

\end{document}